\documentclass[a4paper,12pt]{article}
\usepackage{enumitem}
\usepackage{calc}

\usepackage{lmodern} 

\usepackage{graphicx} 

\usepackage{amsmath, accents}
\usepackage{amssymb}
\usepackage{amsthm}
\usepackage{amscd}
\usepackage{mathrsfs}

\usepackage{yfonts}

\usepackage{marvosym}

\newtheorem{theorem}			     {Theorem} [section]
\newtheorem{proposition}[theorem]	 {Proposition}

\theoremstyle{definition}

\newtheorem{remark} {Remark}

\newcommand{\C}{\mathbb{C}}
\newcommand{\R}{\mathbb{R}}

\newcommand{\Or}{\mathcal{O}}

\newcommand{\K}{\mathcal{K}}

\newcommand{\tr}{\textrm{tr}}

\newcommand{\Ai}{{\rm Ai}}

\usepackage{tikz}
\usetikzlibrary{arrows}
\usetikzlibrary{decorations.pathmorphing}
\usetikzlibrary{decorations.markings}
\usetikzlibrary{patterns}
\usetikzlibrary{automata}
\usetikzlibrary{positioning}
\usepackage{tikz-cd}

\usepackage{pgfplots}

\numberwithin{equation}{section}

\title{The generating function for the Airy point process and a system of coupled Painlev\'e II equations}
\author{Tom Claeys\footnote{Institut de Recherche en Math\'ematique et Physique,  Universit\'e
catholique de Louvain, Chemin du Cyclotron 2, B-1348
Louvain-La-Neuve, BELGIUM} \  and Antoine Doeraene\footnotemark[\value{footnote}]}
\date{}

\tikzset{->-/.style={decoration={
  markings,
  mark=at position #1 with {\arrow{>}}},postaction={decorate}}}
\tikzset{-<-/.style={decoration={
  markings,
  mark=at position #1 with {\arrow{<}}},postaction={decorate}}}

\begin{document}

\maketitle

\begin{abstract}
For a wide class of Hermitian random matrices, the limit distribution of the eigenvalues close to the largest one is governed by the Airy point process. 
In such ensembles, the limit distribution of the $k$-th largest eigenvalue is given in terms of the Airy kernel Fredholm determinant or in terms of Tracy-Widom formulas involving solutions of the Painlev\'e II equation. Limit distributions for quantities involving two or more near-extreme eigenvalues, such as the gap between the $k$-th and the $\ell$-th largest eigenvalue or the sum of the $k$ largest eigenvalues, can be expressed in terms of Fredholm determinants of an Airy kernel with several discontinuities. We establish simple Tracy-Widom type expressions for these Fredholm determinants, which involve solutions to systems of coupled Painlev\'e II equations, and we investigate the asymptotic behavior of these solutions.
\end{abstract}
\section{Introduction}

The Airy point process or Airy ensemble \cite{Borodin, Johansson, PraehoferSpohn} is a determinantal point process which describes, among others, the limit distribution of the largest eigenvalues in many random matrix ensembles as the size of the matrices tends to infinity.
It also appears as limiting process in various other models for repulsive particles, such as random tilings, non-intersecting Brownian paths, and random partitions or Young diagrams with respect to the Plancherel measure, see e.g.\ \cite{Johansson2} for an overview. 
It is a probability distribution on locally finite configurations of real points, characterized by the fact that its $k$-point correlation functions $\rho_k$ take the form 
\[\rho_k(x_1,...,x_k) = \det \left( K^{\Ai}(x_j,x_\ell) \right)_{j,\ell = 1}^k,\]
where
$K^{\Ai}$ is the Airy kernel
\begin{equation}K^\Ai(u, v) = \frac{ \Ai(u) \Ai'(v) - \Ai'(u) \Ai(v) }{ u - v }.\end{equation}
Viewed as an integral kernel operator, $K^{\Ai}$ is trace-class when acting on bounded real intervals or unbounded intervals of the form $(x,+\infty)$.
 We denote the random points in the process by 
\[\zeta_1>\zeta_2>\zeta_3>\ldots,\]
which we can do
since there is almost surely a largest point and since the points are almost surely distinct. Given a random point configuration $\zeta_1,\zeta_2,\ldots$ and a Borel set $A \subseteq \R$, we write $n_A$ for the \emph{occupancy number} of $A$, i.e., the (random) number of points in $A$.

For general determinantal point processes, expectations of the form $\mathbb E\left(\prod_{j = 1}^k s_j^{n_{A_j}}\right)$ for $s_1,\ldots, s_k\in\mathbb C$ and disjoint sets $A_1,\ldots, A_k$ can be expressed as Fredholm determinants. In the case of the Airy point process, we have
\cite[Theorem 2]{Soshnikov2000}
\begin{equation}
\label{eq:generatingasfredholm}
\mathbb E\left( \prod_{j = 1}^k s_j^{n_{A_j}} \right) = \det\left( 1 - \chi_{\cup_j A_j}\sum_{j=1}^k (1 - s_j) K^{\Ai} \chi_{A_j} \right),
\end{equation}
where $K^{\Ai}$ is the integral operator associated to the Airy kernel, and $\chi_A$ is the projection operator from $L^2(\R)$ to $L^2(A)$, or the characteristic function of $A$ at the level of the kernels. We note that this is an entire function of $s_1,...,s_k$.

\paragraph{Generating function.}

We will be interested in the generating function \eqref{eq:generatingasfredholm} of the occupancy numbers in the special case where the sets $A_1,...,A_k$ take the form
\begin{equation}\label{def Aj}
A_j=(x_j,x_{j-1}),\qquad +\infty=:x_0 > x_1 > ... > x_k >-\infty, 
\end{equation}
and where we take $s_1,...,s_k \in [0,1]$. We denote
\begin{equation}
\label{eq:AiryGeneratingFunction}
F(\vec x;\vec s)=F(x_1,...,x_{k};s_1,\ldots, s_k) := \det\left( 1 - \chi_{(x_k, +\infty)} \sum_{j = 1}^k (1 - s_j) K^\Ai \chi_{(x_j, x_{j-1})} \right).
\end{equation}

For $k=1$, the determinant $F(x_1;s_1)$ generates the individual distributions of the largest particles $\zeta_1, \zeta_2,\ldots$ in the Airy point process. Indeed, we have
\begin{equation}\mathbb P(\zeta_1 < x)=\mathbb E(s^{n_{(x,+\infty)}}) \big|_{s = 0}=F(x;0),\end{equation}
and for the $\ell$-th largest particle,
\begin{equation}\label{probk1}\mathbb P(\zeta_\ell < x) = \mathbb P\left( n_{(x,+\infty)} < \ell \right) = \sum_{j = 0}^{\ell-1}\mathbb P\left( n_{(x,+\infty)} = j \right)=\sum_{j = 0}^{\ell-1} \frac{1}{j!} \left.\frac{d^j}{ds^j} F(x;s)\right|_{s = 0},\end{equation}
since
\[F(x;s)=\mathbb E\left( s^{n_{(x,+\infty)}} \right) = \sum_{j = 0}^{+\infty} \mathbb P\left(n_{(x,+\infty)} = j \right) s^j.\]
Tracy and Widom \cite{TracyWidom} showed that the Fredholm determinant $F(x;s)$ can be expressed in terms of a solution to the Painlev\'e II equation: for $0 \leq s < 1$, $x \in \R$, we have
\begin{equation}
\label{eq:tracywidomdistribution}
F(x;s) = \exp\left( - \int_x^{+\infty} (\xi - x) q^2(\xi;s) d\xi \right),
\end{equation}
with $q(\xi;s)$ the solution to the homogeneous Painlev\'e II equation
\begin{equation}\label{PII}q'' = \xi q + 2q^3\end{equation} which is characterized by the following asymptotic behavior at $+\infty$,
\begin{equation}\label{eq:PIIASasymptotics}q(\xi;s) = \sqrt{1-s}\,\Ai(\xi)(1+o(1)), \qquad \xi \to +\infty.\end{equation}
These solutions $q(\xi;s)$ are known as the Ablowitz-Segur solutions \cite{AblowitzSegur} of Painlev\'e II for $0 < s < 1$ and as the Hastings-McLeod solution \cite{HastingsMcLeod} if $s=0$.

Similarly as for $\mathbb P(\zeta_\ell < x)$, we can express the joint probability of two particles $\zeta_{m_1}> \zeta_{m_2}$ with $m_1<m_2$ as
\begin{multline}\label{probk2}\mathbb P(\zeta_{m_1} < x_1, \zeta_{m_2} < x_2) = \sum_{\substack{j_1 < m_1\\j_1+j_2 < m_2}} \mathbb P\Big( n_{(x_1,+\infty)} = j_1, n_{(x_2,x_1)} = j_2 \Big)\\
=\sum_{\substack{j_1 < m_1\\j_1+j_2 < m_2}} \frac{1}{j_1!j_2!} \left. \frac{\partial^{j_1+j_2}}{\partial s_1^{j_1} \partial s_2^{j_2}} F(x_1,x_2;s_1,s_2) \right|_{s_1 = s_2 = 0}.\end{multline}
More generally, the joint distribution of $k$ particles $\zeta_{m_1}>\ldots> \zeta_{m_k}$ with $m_1<\ldots<m_k$ is given by (see e.g.\ \cite{BaikDeiftRains})
\begin{multline}\label{probgenk}\mathbb P\left(\cap_{j=1}^k\left(\zeta_{m_j} < x_j\right)\right) = \sum \mathbb P\Big(\cap_{j=1}^k\left( n_{A_j} = m_j\right)\Big)\\
=\sum \frac{1}{j_1!j_2!\ldots j_k!} \left. \frac{\partial^{j_1+j_2+\ldots j_k}}{\partial s_1^{j_1} \partial s_2^{j_2}\ldots \partial s_k^{j_k}} F(\vec x;\vec s)\right|_{\vec s = 0},\end{multline}
where $A_j$ is as in \eqref{def Aj}, and where the sum is taken over all indices $j_1,\ldots, j_k$ such that 
\begin{equation}
j_1<m_1,\quad j_1+j_2<m_2,\quad  \ldots\quad \sum_{i=1}^k j_i<m_k.
\end{equation}

A Hamiltonian and integrable structure associated to $F(\vec x;\vec s)$ has been established in \cite{HarnadTracyWidom}, and it was linked to a Lie algebra, see also \cite{AdlervanMoerbeke} for a similar approach in a more general context. Given the integrable structure, it is natural to ask whether an explicit formula for $F(\vec x;\vec s)$ in terms of solutions to Painlev\'e-type equations exists.

\paragraph{Tracy-Widom formula.}

The main result of the present paper is a Tracy-Widom type formula for the multi-interval Airy kernel Fredholm determinant $F(x_1,\ldots, x_k;s_1,\ldots, s_k)$ for general $k>1$, similar to \eqref{eq:tracywidomdistribution}.

\begin{theorem}
\label{thm:fredholmdetrep}
Let $s_1, ..., s_k \in [0,1]$ be such that $s_j \neq s_{j+1}$ for $j = 1, \ldots , k$ with $s_{k+1}=1$. Let $x_1,\ldots, x_k$ be as in \eqref{def Aj}, and let $F(\vec x;\vec s)$ be defined by \eqref{eq:AiryGeneratingFunction}. We have
\begin{equation}
\label{eq:theoremformula}
F(\vec x;\vec s) = \prod_{j=1}^k\exp\left( -\int_{0}^{+\infty} \xi u_j(\xi;\vec x,\vec s)^2 d\xi \right),
\end{equation}
where $u_1(\xi;\vec x, \vec s),...,u_k(\xi;\vec x, \vec s)$ satisfy the following system of ordinary differential equations,
\begin{equation}
\label{eq:theoremsystem}
\left\{
\begin{array}{l}
u_1'' = (\xi +x_1) u_1 + 2 u_1 \sum_{j=1}^ku_j^2 \\
u_2'' = (\xi +x_2) u_2 + 2 u_2 \sum_{j=1}^ku_j^2 \\
\vdots \\
u_k'' = (\xi +x_k) u_k + 2 u_k \sum_{j=1}^ku_j^2 \\
\end{array}
\right.
\end{equation} and have the following behavior at $+\infty$,
\begin{equation}\label{eq:asujAiry}u_j(\xi;\vec x,\vec s) = \sqrt{s_{j+1}-s_j} \Ai(\xi+x_j)(1+o(1)),\qquad {\rm as}\,\, \xi\to +\infty,\end{equation}
where we write $s_{k+1}=1$.
If $s_{j+1}>s_{j}$, $u_j(\xi;\vec x,\vec s)$ is real-valued for real $\xi$; if $s_{j+1}<s_{j}$, $u_j(\xi;\vec x,\vec s)$ is purely imaginary for real $\xi$.
\end{theorem}

\begin{remark}
If $s_1<s_2<\ldots < s_k$, all $u_j$'s are real, and then \eqref{eq:theoremformula} can be written as  
\begin{equation}
\label{eq:theoremformula-norm}
F(\vec x;\vec s) = \exp\left( -\int_{0}^{+\infty} \xi\|\vec u(\xi;\vec x,\vec s)\|^2 d\xi \right),
\end{equation}
where
\begin{equation}\label{system-norm}
u_j''=(\xi+x_j)u_j  + 2 u_j\|\vec u\|^2,\qquad j=1,\ldots, k,
\end{equation}
with $\|\vec u\|$ the $2$-norm of the vector $\vec u=(u_1,\ldots, u_k)$.
\end{remark}
\begin{remark}
For $k=1$, the system of ODEs is simply the (shifted) Painlev\'e II equation
\[u_1''=(\xi+x_1)u_1+2u_1^3,\]
and then $q(\xi;s)=u_1(\xi-x_1;x_1,s)$ is the Ablowitz-Segur (for $s\neq 0$) or Hastings-McLeod solution (for $s=0$) of Painlev\'e II. We then easily recover the classical Tracy-Widom formula \eqref{eq:tracywidomdistribution}.
For $k>1$, \eqref{eq:theoremsystem} can be seen as a system of Painlev\'e II equations coupled by the last term at the right hand side.
We will show in Section \ref{section:Lax} that it is the compatibility condition of a Lax pair of size $2\times 2$ with $k$ regular singularities and one irregular singularity at infinity, see \eqref{eq:Laxpair} and \eqref{eq:MatrixAGeneralFormula} below.
\end{remark}

\begin{remark}
In the case where $s_1<\ldots<s_k$, the system \eqref{eq:theoremsystem} of coupled Painlev\'e II equations 
is a {\em traveling wave reduction} of the integrable system of PDEs
\begin{equation}\label{VNLS} i\vec q_t=\vec q_{\xi\xi} - \xi\vec q - 2\vec q\|\vec q\|^2,\qquad \vec q=(q_1,\ldots, q_k),\end{equation}
which is known as the spatially inhomogeneous defocusing {\em vector nonlinear Schr\"odinger equation} with linear potential, see \cite[Equations (2a)--(2b)]{BPB} and also \cite{Hone, VMC} for similar reductions.
For $k=2$, the above system of PDEs is known as the Manakov system \cite{Manakov}.
For traveling wave solutions to \eqref{VNLS} of the form 
\[q_j(\xi;t)=u_j(\xi)e^{-ix_j t},\qquad j=1,\ldots, k,\]
\eqref{VNLS} is easily seen to be equivalent to the system of coupled Painlev\'e II equations \eqref{system-norm}. Also for  $k=2$, this system appeared in Hamiltonian form in a study by Sasano \cite{Sasano} of fourth order extensions of the Painlev\'e II equation.
\end{remark}

\begin{remark}
Theorem \ref{thm:fredholmdetrep} establishes existence of a solution to the system \eqref{eq:theoremsystem} with boundary conditions \eqref{eq:asujAiry}, but not uniqueness. The existence is a consequence of the fact that we will construct the functions $u_j(\xi;\vec x,\vec s)$ explicitly in terms of the solution of a Riemann-Hilbert (RH) problem, and use this representation to show that they satisfy the required boundary conditions. RH characterizations of solutions of Painlev\'e type equations are in general very useful in the study of asymptotics, but are not effective to show uniqueness of solutions with a specific asymptotic behavior.
\end{remark}

Theorem \ref{thm:fredholmdetrep} reveals a connection between the Fredholm determinant $F(\vec x;\vec s)$ and an integrable system of differential equations. One may hope that this relation will help to understand asymptotic properties of the Fredholm determinant $F(\vec x;\vec s)$ in various limits, as it did for $k=1$ in the past. Indeed, \eqref{eq:tracywidomdistribution} together with the asymptotics of the Painlev\'e II solutions $q(\xi;s)$ as $\xi\to -\infty$ provided a simple way to derive  $x\to 
-\infty$ asymptotics for the Fredholm determinant $F(x;s)$, up to a multiplicative constant. In this perspective, it is a natural first step to understand the asymptotic behavior of the functions $u_j(\xi;\vec x,\vec s)$, in particular in limits where the system of $k$ coupled Painlev\'e II equations gets reduced to a system of $k-1$ coupled Painlev\'e II equations.

\paragraph{Asymptotics for the solutions of the system of coupled Painlev\'e II equations.}

The vector solution $\vec u(\xi;\vec x, \vec s)$ exhibits several interesting degenerate cases.
If consecutive values of the parameters $s_{j}, s_{j+1}$ or $x_j, x_{j+1}$ coalesce, the system of $k$ equations simplifies and it formally reduces to the system of $k-1$ coupled Painlev\'e II equations. 
More precisely, if two $s$-values $s_{j}$ and $s_{j+1}$ approach each other, then the component $u_j$ will become small. If we delete the $j$-th equation from the system \eqref{eq:theoremsystem}, the remaining system is again of the same form, but with $k$ reduced by $1$.
If we consecutively let all $s_j$'s for $j=1,\ldots, k$ approach each other, the number of equations in the system gets reduced at each step, and finally asymptotics can be expressed in terms of the Ablowitz-Segur or Hastings-McLeod solutions of the Painlev\'e II equation.
A similar phenomenon takes place if two $x$-values $x_{j+1}$ and $x_j$ coalesce, or if $x_1\to +\infty$. 

Before stating these results in detail, we introduce the following notation: given a vector $\vec v=(v_1,v_2,\ldots, v_k)$, we denote by $\vec v^{[j]}=(v_1,\ldots, v_{j-1}, v_{j+1}, \ldots, v_k)$, for $j=1,\ldots, k$, the vector $\vec v$ without its $j$-th component.

\begin{theorem}\label{thm: asymptotics}
Let $s_1, ..., s_k \in [0,1]$ be such that $s_j \neq s_{j+1}$ for $j = 1, \ldots , k$ with $s_{k+1}=1$, and let $x_1,\ldots, x_k$ be as in \eqref{def Aj}. Let $u_1,\ldots, u_k$ be the solutions appearing in Theorem \ref{thm:fredholmdetrep} of the system of equations \eqref{eq:theoremsystem}. We have the following asymptotic results, which are uniform in $\xi>M$ for any $M\in\mathbb R$.
\begin{enumerate}
\item Let $j\in\{1,\ldots, k\}$. As $s_{j+1}- s_{j}\to 0$ with $s_{j-1}\neq s_{j+1}$ (if $j\neq 1$), we have
\begin{align}
&u_j(\xi;\vec x,\vec s)=\Or\left(|s_j-s_{j+1}|^{1/2}\right),\\
&u_\ell^2(\xi;\vec x,\vec s)= u_\ell^2(\xi;\vec x^{[j]},\vec s^{[j]})+\Or\left(|s_j-s_{j+1}|^{1/2}\right),& \ell<j,\\
&u_\ell^2(\xi;\vec x,\vec s)= u_{\ell-1}^2(\xi;\vec x^{[j]},\vec s^{[j]})+\Or\left(|s_j-s_{j+1}|^{1/2}\right), & \ell>j.
\end{align}
\item Let $j\in\{2,\ldots, k\}$. As $x_{j-1}-x_j\to 0$ with $s_{j-1}\neq s_{j+1}$, we have
\begin{align}
&u_{j-1}^2(\xi;\vec x,\vec s)+u_{j}^2(\xi;\vec x,\vec s)=u_{j-1}^2(\xi;\vec x^{[j]},\vec s^{[j]})+\Or\left(x_{j-1}-x_{j}\right),\\
&u_\ell^2(\xi;\vec x,\vec s)= u_\ell^2(\xi;\vec x^{[j]},\vec s^{[j]})+\Or\left(x_{j-1}-x_{j}\right),& \ell<j-1,\\
&u_\ell^2(\xi;\vec x,\vec s)= u_{\ell-1}^2(\xi;\vec x^{[j]},\vec s^{[j]})+\Or\left(x_{j-1}-x_{j}\right), & \ell>j.
\end{align}
\item If $x_1\to +\infty$, we have
\begin{align}
&u_{1}(\xi;\vec x,\vec s)\sim \sqrt{s_2-s_1}\Ai(\xi+x_1),\\
&u_\ell^2(\xi;\vec x,\vec s)= u_{\ell-1}^2(\xi;\vec x^{[1]},\vec s^{[1]})+\Or\left(x_1^{-1/2}\right), & \ell>1.
\end{align}
\end{enumerate}
\end{theorem}
\begin{remark}
For $k=2$, the above results show that, for $s_2>s_1$, the function $\|\vec u\|=\sqrt{u_1^2+u_2^2}$ reduces to Ablowitz-Segur or Hastings-McLeod solutions to Painlev\'e II in either of the limits $x_1\to x_2$, $x_1\to +\infty$, $s_2\to s_1$, $s_2\to 1$.
We indeed have
\begin{equation}
\|\vec u(\xi;x_1,x_2,s_1,s_2)\|\to
\begin{cases}
 q(\xi+x_2;s_1),& \mbox{as }s_2\to s_1,\\
 q(\xi+x_1;s_1),& \mbox{as }s_2\to 1.
\end{cases}
\end{equation}
In other words, if we let $s_2$ increase from $s_1$ to $1$, $\|\vec u(\xi;x_1,x_2,s_1,s_2)\|$ shows a transition from the Painlev\'e II solution $q(\xi+x_2;s_1)$ to its shifted version $q(\xi+x_1;s_1)$.

Similarly, if we let $x_1$ increase from $x_2$ to $+\infty$, a cross-over between the Painlev\'e II solutions $q(\xi+x_2, s_1)$ and $q(\xi+x_2, s_2)$ takes place:
\begin{equation}
\|\vec u(\xi;x_1,x_2,s_1,s_2)\|\to
\begin{cases}
 q(\xi+x_2;s_1),& \mbox{as }x_1\to x_2,\\
 q(\xi+x_2;s_2),& \mbox{as }x_1\to +\infty.
\end{cases}
\end{equation}
Here, the increase of $x_1$ does not cause a shift in the Painlev\'e variable $\xi$, but in the parameter $s$ which specifies the boundary condition of the Painlev\'e II solution.

On the level of the Fredholm determinant $F(\vec x;\vec s)$ and on the level of an associated RH problem, these shifts were already observed in 2001 by Baik, Deift, and Rains \cite[Section 6]{BaikDeiftRains}, who noticed that these phenomena are reminiscent of the behavior of superposed soliton solutions to nonlinear wave equations after they collided. The terminology {\em Painlev\'e-tons}, or {\em multi-Painlev\'e} functions, was suggested in \cite{BaikDeiftRains}.
\end{remark}

\begin{remark}
In principle, the asymptotic analysis in Section \ref{section:RH} allows with some more effort to compute further subleading terms in the asymptotic expansions from Theorem \ref{thm: asymptotics}, expressed in terms of the solution to the RH problem for $\Psi$ stated below. However, it is not clear that this would lead to simple explicit expressions for these subleading terms.
\end{remark}

\paragraph{Outline.}
In Section \ref{section:examples}, we will describe some concrete examples of probability distributions which one can compute using the Fredholm determinants $F(\vec x;\vec s)$, and we express those distributions explicitly in terms of the solutions $\vec u(\xi;\vec x,\vec s)$ of the system of coupled Painlev\'e II equations. In Section \ref{section:Lax}, we will express logarithmic derivatives of the Fredholm determinants in terms of a RH problem, and we will derive an associated Lax pair. By manipulating the Lax pair in a convenient way, we will prove Theorem \ref{thm:fredholmdetrep}, except for the asymptotic formula \eqref{eq:asujAiry}.
In Section \ref{section:RH}, we will analyze the RH problem associated to the Fredholm determinants and the functions $u_j$ asympotically, and this will allow us to prove \eqref{eq:asujAiry} and Theorem \ref{thm: asymptotics}.

\section{Examples and applications}\label{section:examples}

\paragraph{Near-extreme GUE eigenvalues.} The prototype example of a model in which the Airy point process appears as a limiting process is the GUE.
The eigenvalues $\lambda_1>\lambda_2>\ldots>\lambda_n$ of an $n\times n$ GUE matrix follow a determinantal point process with kernel $K_n$ built out of Hermite polynomials,
\begin{equation}\label{eq:kernel GUE}
K_n^{\rm GUE}(x,y)= e^{-\frac{1}{4}(x^2 + y^2)} \sum_{j = 0}^{n-1} p_j(x)p_j(y),\end{equation}
where $(p_j)_{j=0,1,\ldots}$ is the sequence of normalized Hermite polynomials, orthogonal with respect to the weight $e^{-x^2/2}$ on $\mathbb R$.

After the re-scaling
\begin{equation}\label{eq:scalingxlambda}x_j=n^{1/6}(\lambda_j-2\sqrt{n}),
\end{equation} this kernel converges to the Airy kernel,
\begin{equation}\label{eq:scaling GUE}
\lim_{n\to\infty}n^{-1/6}K_n^{\rm GUE}(2\sqrt{n}+x n^{-1/6},2\sqrt{n}+y n^{-1/6})= K^{\rm Ai}(x,y),\end{equation}
uniformly for $x,y\in [M,+\infty)$ for any $M\in\mathbb R$. This implies also trace-norm convergence of the associated operators when acting on bounded intervals or unbounded intervals of the form $[M,+\infty)$.
Therefore, with $\vec x$ and $\vec\lambda$ related as in \eqref{eq:scalingxlambda}, we have the limit \begin{equation}\label{eq:FredholmGUE}F_n^{\rm GUE}(\vec\lambda;\vec s):=\det\left( 1 - \chi_{(\lambda_k, +\infty)} \sum_{j = 1}^k (1 - s_j) K_n \chi_{(\lambda_j, \lambda_{j-1})} \right)\rightarrow F(\vec x;\vec s),\qquad n\to\infty,\end{equation}
see e.g.\ \cite{Soshnikov2000}. 
It is worth noting that $F_n^{\rm GUE}$ can also be expressed as a ratio of Hankel determinants with respect to a discontinuous Gaussian weight,
\begin{equation}\label{eq:Hankel}
F_n^{\rm GUE}(\vec\lambda;\vec s)=\frac{\det\left(\int_{\mathbb R}x^{\ell+m}e^{-\frac{x^2}{2}}\sum_{j=1}^{k+1} s_j \chi_{(x_{j},x_{j-1})}(x)dx\right)_{\ell,m=0}^{n-1}}{\det\left(\int_{\mathbb R}x^{\ell+m}e^{-\frac{x^2}{2}}dx\right)_{\ell,m=0}^{n-1}},
\end{equation}
with $x_0=+\infty$, $x_{k+1}=-\infty$, $s_{k+1}=1$.
Since the denominator is a Selberg integral which can be evaluated explicitly, Theorem \ref{thm:fredholmdetrep} together with \eqref{eq:FredholmGUE} also implies an asymptotic expansion for the Hankel determinant in the numerator of \eqref{eq:Hankel}, which is a generalization of the asymptotic expansion from \cite{BogatskiyClaeysIts} in the case of $1$ jump discontinuity.

The joint probability distribution of $k$ near-extreme eigenvalues in the GUE can be expressed exactly in terms of $F_n^{\rm GUE}(\vec\lambda;\vec s)$ in a similar way as in formula \eqref{probgenk} for the Airy point process. In particular, gap probabilities near the edge, the distribution of spacings between the $k$-th and the $\ell$-th largest eigenvalue, the sum of the $k$ largest eigenvalues, and the distribution of truncated linear statistics of the form $\sum_{j=1}^k f(\lambda_j)$ (as studied recently in \cite{GMT}) can be expressed identically in terms of $F_n(\vec\lambda;\vec s)$. In the large $n$ limit, after the re-scaling \eqref{eq:scaling GUE}, limit distributions arise which are given in terms of $F(\vec x;\vec s)$.

\paragraph{Gap probabilities.}
The probability to find no GUE eigenvalues $\lambda_j$ in a scaled interval $\left(2\sqrt{n}+x_2 n^{-1/6}, 2\sqrt{n}+x_1 n^{-1/6}\right)$ near the edge of the spectrum converges as $n\to\infty$ to the probability of having no
particles $\zeta_j$ of the Airy point process in a finite interval $(x_2,x_1)$, and this is given by
\begin{equation}\label{gap probability}
\mathbb P(n_{(x_2,x_1)}=0)=F(x_1,x_2;1,0)=\exp\left(\int_{0}^{+\infty} \xi |u_1(\xi)|^2 d\xi \right)\exp\left( -\int_{0}^{+\infty} \xi |u_2(\xi)|^2 d\xi \right)
\end{equation}
with
\begin{equation}
u_j''(x)=(x+x_j) u_j(x) +2u_j(x)(u_1(x)^2 + u_2(x)^2),\quad j=1,2\label{eq:systemk2}
\end{equation}
and
\[u_{j}(x)^2\sim (-1)^j {\rm Ai}(x+x_j)^2,\quad x\to +\infty,\quad j=1,2.\]
As $x_1\to +\infty$, it is obvious that this gap probability converges to the largest eigenvalue distribution $F(x_2;0)$, which can also be seen easily from Theorem \ref{thm: asymptotics}. Setting $k \geq 2$ in Theorem \ref{thm:fredholmdetrep}, we can find expressions for gap probabilities on any finite union of intervals.

\paragraph{Thinning and a conditional largest eigenvalue distribution.}

The thinned Airy point process is the process obtained by removing each particle $\zeta_1,\zeta_2,\ldots$ independently with a given probability $s\in (0,1)$. The parameter $s$ can be seen as a measure for repulsion: $s=1$ corresponds to a repulsive point process and the limit $s\to 0$ to a Poisson process. In \cite{Bohigas-deCarvalho-Pato, Bothner-Buckingham}, the largest particle distribution in this process was studied and a transition between the Tracy-Widom distribution and the Gumbel distribution was observed as $s$ tends to $1$.
One can interpret the removed particles in this process as unobserved and the remaining ones as observed \cite{BohigasPato2}.
It is then natural to ask what the distribution of the largest particle $\zeta_1$ of the Airy point process is, given information about the position of the largest observed particle, which we denote by $\xi_1$ and which is equal in distribution to $\zeta_{Y}$, with $Y$ a geometric random variable defined by $\mathbb P(Y=k)=(1-s)s^{k-1}$.
More precisely, we consider the conditional probability distribution of $\zeta_1$, conditioning on the event that the largest observed particle $\xi_1$ is less than a given value.
We note in this context that the observed particles also form a determinantal point process, with correlation kernel $(1-s)K^{\rm Ai}$ \cite{LavancierMollerRubak}, and therefore the distribution of $\xi_1$ is given by $\mathbb P(\xi_1<x_2)=F(x_2;s)$.

For $x_1<x_2$, the conditional probability that $\zeta_1<x_1$, given that $\xi_1<x_2$ is clearly equal to the ratio $F(x_1;0)/F(x_2;s)$.
For $x_1>x_2$, it can be expressed in terms of $F(x_1,x_2;0,s)$ as follows,
\begin{multline*}
\mathbb P\left(\zeta_1<x_1| \xi_1<x_2\right)=\frac{\mathbb P\left(\zeta_1<x_1,  \xi_1<x_2\right)}{\mathbb P\left(\xi_1<x_2\right)}=\frac{\sum_{j=0}^\infty s^j\mathbb P\left(n_{(x_1,+\infty)}=0,n_{(x_2,x_1)}=j\right)}{\mathbb P\left(\xi_1<x_2\right)}
\\
=\frac{\mathbb E\left(s^{n_{(x_2,x_1)}}\right)}{\mathbb P\left(\xi_1<x_2\right)}=\frac{F(x_1,x_2;0,s)}{F(x_2;s)}.
\end{multline*}
Using \eqref{eq:tracywidomdistribution} and \eqref{eq:theoremformula}, we can write this as
\begin{multline}\label{eq:conditional}
\mathbb P\left(\zeta_1<x_1|\xi_1<x_2\right)
=\exp\left(\int_{0}^{+\infty} \xi q(\xi+x_2;s)^2 d\xi \right)\\
\times \exp\left(-\int_{0}^{+\infty} \xi u_1(\xi;x_1,x_2,0,s)^2 d\xi \right)\exp\left( -\int_{0}^{+\infty} \xi u_2(\xi;x_1,x_2,0,s)^2 d\xi \right),
\end{multline} where $q$ is the Ablowitz-Segur solution of Painlev\'e II, and where $u_1$ and $u_2$ solve the system \eqref{eq:systemk2} with asymptotic behavior
\begin{align*}&u_{1}(\xi)^2\sim  s {\rm Ai}(\xi+x_1)^2,&& \xi\to +\infty\\
&u_{2}(\xi)^2\sim (1-s) {\rm Ai}(\xi+x_2)^2,&& \xi\to +\infty.
\end{align*}

Simultaneously with our work, a determinant closely related to $F(x_1,x_2;0,s)$ but with an extra spectral singularity was studied by Xu and Dai in \cite[Theorem 2]{XuDai}, where they obtained a Tracy-Widom formula which is equivalent to ours if the spectral singularity is absent.

As $x_1\to x_2$, one can use Theorem \ref{thm: asymptotics} to confirm that \eqref{eq:conditional} tends to $\frac{P(\zeta_1 < x_1)}{P(\xi_1 < x_1)}=F(x_1;0)/F(x_2;s)$.
As $s\to 0$, Theorem \ref{thm: asymptotics} implies that \eqref{eq:conditional} converges to $1$, which is natural since almost all eigenvalues are observed. If $s\to 1$ on the other hand, Theorem \ref{thm: asymptotics} implies that \eqref{eq:conditional} converges to the Tracy-Widom distribution $F(x_1;0)$, which is again what could be expected: since almost no eigenvalues are observed, we condition on a very likely event, and the conditional largest particle distribution converges to the unconditioned largest particle distribution.

The thinned Airy point process arises not only as the large $n$ limit of the thinned GUE, but also as limit point process in domino tilings of Aztec diamonds with different weights for horizontal and vertical dominoes \cite{ChhitaJohanssonYoung}.

\paragraph{Distribution of the spacing between the largest two particles.}
The distribution of the spacing between the largest two eigenvalues of a GUE matrix has attracted a lot of interest recently, see \cite{PerretSchehr, WBF} and also \cite{DeiftTrogdon} where it was shown that this distribution is directly related to the distribution of the first halting time of the Toda eigenvalue algorithm applied to random matrices.
The limit distribution of this spacing, after re-scaling, is given by $\mathbb P(\zeta_1-\zeta_2>\sigma)$, with $\zeta_1>\zeta_2$ the largest two particles in the Airy point process.
This distribution can be expressed as
\begin{align*}
\mathbb P(\zeta_1-\zeta_2>\sigma)&=\int_{\mathbb R} \frac{\partial}{\partial x_1}\left.\mathbb P(\zeta_1<x_1,\zeta_2<\zeta-\sigma)\right|_{x_1=\zeta}      d\zeta
\\
&=\int_{\mathbb R}\frac{\partial}{\partial x_1} \left.\left(\mathbb P(n_{(x_1,+\infty)}=0, n_{(\zeta-\sigma,\zeta)}=0) + \mathbb P(n_{(x_1,+\infty)}=0, n_{(\zeta-\sigma,\zeta)}=1\right)\right|_{x_1=\zeta}   d\zeta\\
&=\int_{\mathbb R}\frac{\partial^2}{\partial s_2\partial x_1} \left.\left(F(x_1,\zeta-\sigma;0,s_2)\right)\right|_{x_1=\zeta, s_2=0}   d\zeta, 
\end{align*}
by 
\eqref{probk2}.
Using \eqref{eq:theoremformula}, this can be rewritten as
\begin{equation}\label{eq:distrspacing}\mathbb P\left(\zeta_1-\zeta_2>\sigma\right)=\int_{\mathbb R}v(\zeta+\sigma,\zeta)F^{\rm TW}(\zeta;0)d\zeta,\end{equation}
where
\begin{equation}
\label{eq:def v}v(x_1,x_2)=\int_{\mathbb R}\xi\left.\frac{-\partial^2}{\partial s_2 \partial x_1} \left(u_1^2(\xi;x_1,x_2;0,s_2)+u_2^2(\xi;x_1,x_2;0,s_2)\right)\right|_{s_2=0} d\xi,
\end{equation}
and $u_1, u_2$ solve the system of equations \eqref{eq:systemk2} with asymptotic behavior \eqref{eq:asujAiry}.

Different expressions for this distribution were obtained in \cite{PerretSchehr} and \cite{WBF}, in terms of fundamental solutions of the Lax pair associated to the Hastings-McLeod solution of the Painlev\'e II equation. The large and small $\sigma$ asymptotics of the distribution \eqref{eq:distrspacing} are of particular interest and were obtained in \cite{PerretSchehr}. To derive them from our formula, a more detailed understanding of the asymptotic behavior of $v(x_1,x_2)$ would be needed as $x_1-x_2\to+\infty$ or as $x_1-x_2\to 0$.

\paragraph{Plancherel measure and maximal sum of lengths of $k$ disjoint increasing subsequences of a random permutation.}
Another model in which the Airy point process arises naturally consists of random partitions with respect to the Plancherel measure.
Consider the joint distribution of the $k$ first components $\lambda_1,\ldots, \lambda_k$ of a random partition of $n$ or a Young diagram of $n$ boxes following the Plancherel measure. If we set $\zeta_j=n^{-1/6}(\lambda_j-\sqrt{2n})$ , then the joint distribution of $\zeta_1,\ldots, \zeta_k$ converges as $n\to +\infty$ to the joint distribution of the $k$ largest particles in the Airy point process \cite{BDJ, Okounkov, BOO, Johansson}. From the Robinson-Schensed-Knuth correspondence, it follows that the sum of the first $k$ components $S_k:=\lambda_1+\lambda_2+\ldots +\lambda_k$ has the same distribution as the maximal sum of the lengths of $k$ disjoint increasing subsequences of a random permutation (see e.g.\ \cite{BDS, Romik}).

For $k=2$,
a straightforward calculation similar to the one for the spacing leads to 
\begin{align*}
\lim_{n\to\infty}\mathbb P\left(n^{1/6}(S_2-4\sqrt{n})<\sigma\right)&=\mathbb P(\zeta_1+\zeta_2<\sigma)=\int_{\mathbb R} \frac{\partial}{\partial x_1}\left.\mathbb P(\zeta_1<x_1,\zeta_2<\sigma-\zeta)\right|_{x_1=\zeta}      d\zeta
\\
&=\int_{\mathbb R}\frac{\partial^2}{\partial s_2\partial x_1} \left.\left(F(x_1,\sigma-\zeta;0,s_2)\right)\right|_{x_1=\zeta, s_2=0}   d\zeta.
\end{align*}
Using \eqref{eq:theoremformula}, we can write this after a straightforward calculation as
\begin{equation}\label{eq:distrsum}\mathbb P\left(\zeta_1+\zeta_2<\sigma\right)=\int_{\mathbb R}v(\sigma-\zeta,\zeta)F^{\rm TW}(\zeta;0)d\zeta,\end{equation}
with $v$ as defined in \eqref{eq:def v}.

Hence, by \eqref{eq:distrsum}, it follows that
\begin{equation}
\lim_{n\to\infty}\mathbb P\left(n^{1/6}(S_2-4\sqrt{n})<\sigma\right)=\int_{\mathbb R}v(\sigma-\zeta,\zeta)F^{\rm TW}(\zeta;0)d\zeta,
\end{equation}
where $S_2$ is the maximal total length of two disjoint increasing subsequences of a random permutation of $n$. Similar but lengthier formulas can be obtained for general $k$.

\section{Differential identities and Lax pair}\label{section:Lax}

\subsection{Differential identities in terms of RH problem}

In this section, we will relate the Fredholm determinants $F(\vec x;\vec s)$ with the functions $u_j(\xi;\vec x,\vec s)$ via a RH problem. The RH problem depends on parameters $x, s_1, \ldots, s_k$ and $y_1>y_2>\ldots > y_{k-1}>0$, where $y_j$ will be identified with $x_j-x_{k}$, and $x$ with $\xi+x_k$.
We first state an identity which expresses logarithmic derivatives of the Fredholm determinants $F$ in terms of a RH problem.

\begin{proposition}\label{prop: diff id FPsi}
Let $F(\vec x;\vec s)$ be as in \eqref{eq:AiryGeneratingFunction}, with $\vec s$ and $\vec x$ as in Theorem \ref{thm:fredholmdetrep}.
Define $\vec y=(y_1,\ldots, y_{k-1})$ with $y_j=x_j-x_k>0$. We have the differential identities
\begin{equation}
\label{eq:differential-equation}
\frac{\partial}{\partial x_j}\log F(\vec x;\vec s)=\frac{s_{j+1}-s_j}{2\pi i}\lim_{\zeta\to y_j}\left(\Psi^{-1}\Psi_\zeta\right)_{2,1}(\zeta;x=x_k,\vec y,\vec s),\qquad j=1,\ldots, k,
\end{equation}
with $\Psi(\zeta;x,\vec y,\vec s)$ the unique solution of the RH problem below, which depends on parameters $x\in\mathbb R$, $y_{k-1}<y_{k-2}<\ldots<y_1$ and $s_1,\ldots, s_k$, and where $\Psi_\zeta$ is the derivative of $\Psi$ with respect to $\zeta$.
\end{proposition}
\paragraph{RH problem for $\Psi$.}

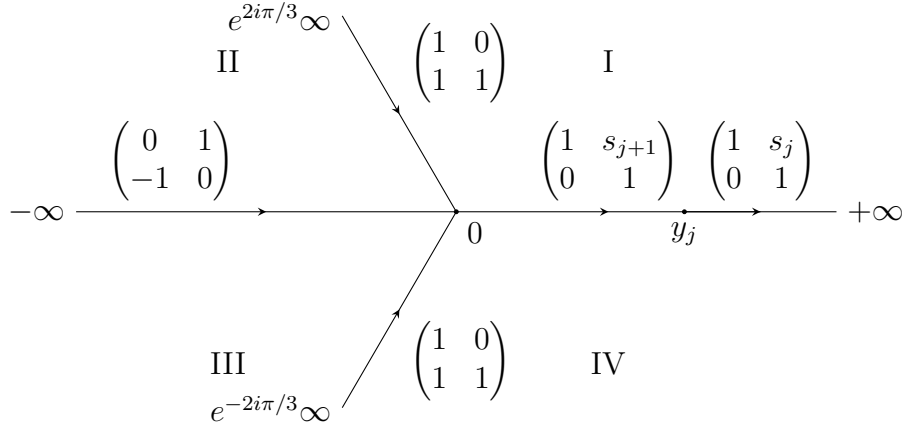
\begin{figure}
\centering
\begin{tikzpicture}[> = stealth]

\draw [->- = .5] (120:3cm) -- node [anchor = south west] {$\begin{pmatrix} 1 & 0 \\ 1 & 1 \end{pmatrix}$} (0,0);
\draw [->- = .5] (-5,0) -- node [near start, anchor = south] {$\begin{pmatrix} 0 & 1 \\ -1 & 0 \end{pmatrix}$} (0,0);
\draw [->- = .5] (-120:3cm) -- node [anchor = north west] {$\begin{pmatrix} 1 & 0 \\ 1 & 1 \end{pmatrix}$} (0,0);
\draw [->- = .5] (0,0) -- node [anchor = south] {$\begin{pmatrix} 1 & s_{j+1} \\ 0 & 1 \end{pmatrix}$} (4,0);
\draw [->- = .5] (3,0) -- node [anchor = south] {$\begin{pmatrix} 1 & s_j \\ 0 & 1 \end{pmatrix}$} (5,0);

\draw (0,0) node [anchor = north west] {$0$};
\draw (3,0) node [anchor = north] {$y_j$};
\draw (120:3cm) node [anchor = east] {$e^{2i\pi/3} \infty$};
\draw (-120:3cm) node [anchor = east] {$e^{-2i\pi/3} \infty$};
\draw (-5,0) node [anchor = east] {$-\infty$};
\draw (5,0) node [anchor = west] {$+\infty$};
\filldraw (0,0) circle (0.025cm);
\filldraw (3,0) circle (0.025cm);

\draw (2,2) node {I};
\draw (-3,2) node {II};
\draw (-3,-2) node {III};
\draw (2,-2) node {IV};

\end{tikzpicture}
\caption{Jump contours for the model RH problem for $\Psi$.}
\label{fig:modelRHcontours}
\end{figure}

\begin{enumerate}[label={(\alph*)}]
\item[(a)] $\Psi : \C \backslash \Gamma \rightarrow \C^{2\times 2}$ is analytic, with
\begin{equation}\label{eq:defGamma}
\Gamma=\mathbb R\cup e^{\pm \frac{2\pi i}{3}} (0,+\infty)
\end{equation}
and $\Gamma$ oriented as in Figure \ref{fig:modelRHcontours}.
\item[(b)] $\Psi(\zeta)$ has continuous boundary values as $\zeta\in\Gamma\backslash \{y_1,\ldots, y_k\}$ is approached from the left ($+$ side) or from the right ($-$ side) and they are related by
\begin{equation*}
\left\{
\begin{array}{ll}
\Psi_+(\zeta) = \Psi_-(\zeta) \begin{pmatrix} 1 & 0 \\ 1 & 1 \end{pmatrix} & \textrm{for } \zeta \in e^{\pm \frac{2\pi i}{3}} (0,+\infty), \\
\Psi_+(\zeta) = \Psi_-(\zeta) \begin{pmatrix} 0 & 1 \\ -1 & 0 \end{pmatrix} & \textrm{for } \zeta \in (-\infty,0), \\
\Psi_+(\zeta) = \Psi_-(\zeta) \begin{pmatrix} 1 & s_j \\ 0 & 1 \end{pmatrix} & \textrm{for } \zeta \in (y_{j}, y_{j-1}),  j=1,\ldots, k,
\end{array}
\right.
\end{equation*}
where we write $y_k=0$, $y_0=+\infty$.
\item[(c)] As $\zeta \rightarrow \infty$, there exist functions $p=p(x;\vec y,\vec s), q=q(x;\vec y,\vec s)$ and $r=r(x;\vec y,\vec s)$ such that $\Psi$ has the asymptotic behavior
\begin{equation}
\label{eq:psiasympinf}
\Psi(\zeta) = \left( I + \frac{1}{\zeta} \begin{pmatrix} q & -i r \\ i p & -q \end{pmatrix} + \Or\left(\frac{1}{\zeta^2}\right) \right) \zeta^{\frac{1}{4} \sigma_3} M^{-1} e^{-(\frac{2}{3}\zeta^{3/2} + x\zeta^{1/2}) \sigma_3},
\end{equation}
where $M = (I + i \sigma_1) / \sqrt{2}$, $\sigma_1 = \begin{pmatrix} 0 & 1 \\ 1 & 0 \end{pmatrix}$ and $\sigma_3 = \begin{pmatrix} 1 & 0 \\ 0 & -1 \end{pmatrix}$.
\item[(d)] $\Psi(\zeta) = \Or( \log(\zeta-y_j) )$ as $\zeta \rightarrow y_j$, $j = 1, ..., k$.
\end{enumerate}

We proceed with the proof of Proposition \ref{prop: diff id FPsi}, which relies on a well-known procedure developed by Its, Izergin, Korepin, and Slavnov (\cite{IIKS}, see also \cite{DeiftItsZhou} in general and \cite{BertolaCafasso, CIK} for applications of this procedure in cases which are similar to ours).

\begin{proof}

Let $K$ be the integral operator kernel
\[K(u,v) = \chi_{(x_k, +\infty)}(u) \sum_{j = 1}^k (1 - s_j) K^\Ai(u,v) \chi_{(x_j, x_{j-1})}(v),\]
and $\K$ the operator it represents. The kernel $K$ is integrable in the sense of Its, Izergin, Korepin, and Slavnov since it can be written in the form
\[K(u,v) = \frac{f^t(u) h(v)}{u - v}\]
with
\begin{equation}
\label{eq:deffh}
f(z) = \begin{pmatrix} \Ai(z) \chi_{(x_k,+\infty)}(z) \\ \Ai'(z) \chi_{(x_k,+\infty)}(z) \end{pmatrix}, \qquad h(z) = \begin{pmatrix} \sum_{j = 1}^k (1 - s_j) \Ai'(z) \chi_{(x_j, x_{j-1})}(z) \\ -\sum_{j = 1}^k (1 - s_j) \Ai(z) \chi_{(x_j, x_{j-1})}(z) \end{pmatrix},
\end{equation}
so that $f^t(u)h(u)=0$.
By \eqref{eq:generatingasfredholm}, and the fact that $s_j \geq 0$ for all $j = 1,...,k$, we have that the operator $I - \K$ is invertible. Using standard properties of trace-class operators, we have for any $j = 1,...,k$,
\begin{equation}
\label{eq:differential-equation-resolvent}
\frac{\partial}{\partial x_j} \ln \det(I - \K) = -\tr \left( (I-\K)^{-1} \frac{\partial \K}{\partial x_j} \right) = \frac{s_{j+1}-s_j}{1-s_j} R(x_j,x_j),
\end{equation}
where $R$ is the integral operator kernel of the resolvent $\mathcal{R}$ of $\K$, defined by the operator relation
\[I + \mathcal{R} = (I-\K)^{-1},\]
and $R(x_j,x_j)$ has to be understood as the limit
\[R(x_j,x_j) = \lim_{x \searrow x_j} R(x,x).\]
Here we suppose that $s_j \neq 1$ for each $j = 1,...,k$. If  $s_j = 1$, then the limit has to be taken from the left instead, and $1-s_j$ has to be replaced by $1-s_{j+1}$ which will be different from $0$, since $s_j \neq s_{j+1}$.

If we define 
\[Y(\zeta)=I-\int_{x_k}^{+\infty}\frac{F(\mu)h^t(\mu)}{\mu-\lambda}d\mu, \qquad F=\begin{pmatrix}(1-\mathcal K)^{-1}f_1\\
(1-\mathcal K)^{-1}f_2\end{pmatrix},\]
then $Y$ satisfies the following RH problem, see \cite{IIKS}.

\paragraph{RH problem for $Y$}
\begin{itemize}
\item[(a)] $Y : \C \setminus [x_k,+\infty) \to \C^{2\times 2}$ is analytic
\item[(b)] $Y$ has continuous boundary values when $\zeta\in(x_k, +\infty) \setminus \{x_1,...,x_k\}$ is approached from the left ($+$ side) or right ($-$ side) and both limits are related via
\[Y_+(\zeta) = Y_-(\zeta)J(\zeta), \qquad J(\zeta) = 1 - 2\pi i f(\zeta)h^t(\zeta).\]
\item[(c)] $Y(\zeta) = I + \Or(1/\zeta)$, as $\zeta \to \infty$,
\item[(d)] $Y(\zeta) = \Or(\log(|\zeta-x_j|))$ as $\zeta \to x_j$, $j = 1, ..., k$.
\end{itemize}
In terms of the solution $Y$, we have (see \cite[Lemmas 2.8 and 2.12]{DeiftItsZhou}),
\begin{equation}
\label{eq:resolvent-integrable}
R(u,v) = \frac{F^t(u)H(v)}{u-v},
\end{equation}
with
\[F(u) = Y_+(u)f(u), \qquad  H(v) = (Y_+^{-1}(v))^th(v).\]
In other words, the integral kernel $R$ is also integrable, and can be expressed explicitly in terms of the unique solution to the above RH problem.

\begin{figure}
\centering
\begin{tikzpicture}[> = stealth]

\draw [->- = .5] (120:3cm) -- (0,0);
\draw [->- = .5] (-5,0) -- (0,0);
\draw [->- = .5] (-120:3cm) -- (0,0);
\draw [->- = .5] (0,0) -- (3,0);
\draw [->- = .5] (3,0) -- (5,0);

\draw (0,0) node [anchor = north west] {$x_k$};
\draw (3,0) node [anchor = north] {$x_1$};
\draw (120:3cm) node [anchor = east] {$e^{2i\pi/3} \infty$};
\draw (-120:3cm) node [anchor = east] {$e^{-2i\pi/3} \infty$};
\draw (-5,0) node [anchor = east] {$-\infty$};
\draw (5,0) node [anchor = west] {$+\infty$};
\filldraw (0,0) circle (0.025cm);
\filldraw (3,0) circle (0.025cm);

\draw (2,1) node {$A$};
\draw (-3,1) node {$B$};
\draw (-3,-1) node {$C$};
\draw (2,-1) node {$D$};

\end{tikzpicture}
\caption{Regions for the definition of $\Phi$.}
\label{fig:XRHcontours}
\end{figure}
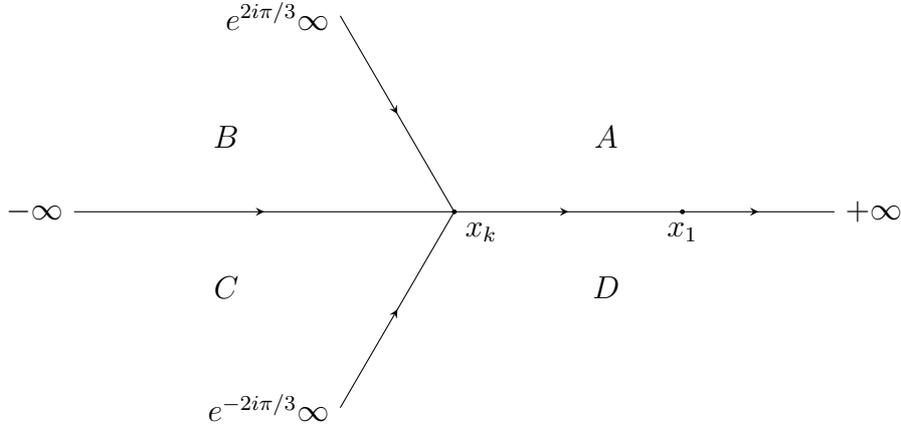

Consider the regions $A$, $B$, $C$ and $D$ in Figure \ref{fig:XRHcontours} and define matrices $\Phi_A, \Phi_B, \Phi_C, \Phi_D$ in terms of the Airy function as follows,
\begin{align}
&\label{def:PhiA}\Phi_A(z) = \sqrt{2\pi} e^{-\pi i / 12} 
\begin{pmatrix} \Ai(z) & \Ai(\omega^2 z) \\ \Ai'(z) & \omega^2 \Ai'(\omega^2 z) \end{pmatrix} e^{-\pi i / 6 \sigma_3} ,\\
&\Phi_D(z) = \sqrt{2\pi} e^{-\pi i / 12} \begin{pmatrix} \Ai(z) & -\omega^2\Ai(\omega z) \\ \Ai'(z) & - \Ai'(\omega z) \end{pmatrix} e^{-\pi i / 6 \sigma_3},\\
&\Phi_B(z) = \Phi_A(z) \begin{pmatrix} 1 & 0 \\ -1 & 1 \end{pmatrix}, \\
&\label{def:PhiC}\Phi_C(z) = \Phi_D(z) \begin{pmatrix} 1 & 0 \\ 1 & 1 \end{pmatrix}
,
\end{align}
with $\omega = e^{\frac{2\pi i}{3}}$, and let 
\begin{equation}\label{def:Phi}
\Phi(z)=\Phi_{*}(z),\qquad \mbox{ for }z \mbox{ in region }*= A,B,C \mbox{ or }D.
\end{equation}
$\Phi$ is the solution to the standard Airy model RH problem (see e.g.\ \cite{DeiftKriecherbauerMcLaughlinVenakidesZhou}): it satisfies the jump relations
\begin{equation*}
\begin{array}{ll}
\Phi_+(z) = \Phi_-(z) \begin{pmatrix} 1 & 0 \\ 1 & 1 \end{pmatrix} & \textrm{for } z \in x_k+ e^{\pm \frac{2\pi i}{3}} (0,+\infty), \\
\Phi_+(z) = \Phi_-(z) \begin{pmatrix} 0 & 1 \\ -1 & 0 \end{pmatrix} & \textrm{for } z \in (-\infty,x_k), \\
\Phi_+(z) = \Phi_-(\zeta) \begin{pmatrix} 1 & 1 \\ 0 & 1 \end{pmatrix} & \textrm{for } z \in (x_k,+\infty),\end{array}
\end{equation*}
and it has the asymptotics
\begin{equation}\label{eq:asPhi} \Phi(z) = z^{-\sigma_3/4} \frac{1}{\sqrt{2}} \begin{pmatrix} 1 & 1 \\ -1 & 1 \end{pmatrix} \left( I + \Or\left( \frac{1}{z^{3/2}} \right) \right) e^{-\pi i \sigma_3 / 4} e^{- \frac{2}{3} z^{3/2} \sigma_3}, \textrm{ as } z \rightarrow \infty.\end{equation}

We define
\[\Psi(z) = \begin{pmatrix} 1 & i \frac{x_k^2}{4} \\ 0 & 1 \end{pmatrix} e^{\frac{i\pi}{4} \sigma_3} \begin{pmatrix} 0 & -1 \\ 1 & 0 \end{pmatrix} Y(z + x_k) \Phi(z + x_k),\]
and it is then straightforward, by the RH conditions for $\Phi$, to verify that $\Psi$ is the (unique) solution to the RH problem for $\Psi$ stated before.
\end{proof}

\begin{remark}
The existence of a solution to the RH  problem for $\Psi$ is a consequence of the above proof. Indeed, we started with a Fredholm determinant $\det(I-\mathcal K)$ in \eqref{eq:deffh} which is non-zero by \eqref{eq:generatingasfredholm}, and the solutions to the RH problems for $Y$ and $\Psi$ are constructed explicitly in terms of the inverse of the operator $I-\mathcal K$, hence those RH problems are solvable. In more general situations, one cannot always rely on a probabilistic interpretation like \eqref{eq:generatingasfredholm}. In such cases, an alternative is to use the vanishing lemma approach. This was done for instance in \cite[Lemma 1]{XuZhao} for $k=1$, and the proof generalizes in a straightforward way to the case of general $k$.
Uniqueness of the solution follows from standard arguments.
\end{remark}

The next proposition relates the RH solution $\Psi$, and in particular the right hand side of the differential identities in Proposition \ref{prop: diff id FPsi}, to the system of differential equations \eqref{eq:theoremsystem}.

\begin{proposition}
\label{prop Psi u}
Let $\Psi(\zeta;x,\vec y,\vec s)$ be the solution of the above RH problem, with $\vec s$ and $\vec y$ as in Proposition \ref{prop: diff id FPsi}. 
If we define
\begin{equation}
u_j^2(\xi;\vec x,\vec s)=
-\frac{s_{j+1}-s_j}{2\pi}\lim_{\zeta\to y_j}\Psi_{2,1}^2(\zeta;\xi+x_k,\vec y,\vec s),\qquad j=1,\ldots, k\label{eq:uintermsofPsi21}
\end{equation}
with the limit $\zeta\to y_j$ taken as $\zeta$ approaches $y_j$ from region I,
then 
\begin{enumerate}
\item $\vec u(\xi;\vec x,\vec s)$ satisfies the system of equations \eqref{eq:theoremsystem},
\item we have the identity
\begin{equation}
u_j^2(x-x_k;\vec x,\vec s)=-\frac{s_{j+1}-s_j}{2\pi i}\frac{\partial}{\partial x}\left(\Psi^{-1}\Psi_\zeta\right)_{2,1}(\zeta=y_j;x,\vec y,\vec s).
\end{equation}
\end{enumerate}
\end{proposition}
The proof of this result is given in the next subsection, and is based on Lax pair techniques.

\subsection{Proof of Proposition \ref{prop Psi u}}

In this subsection, the vectors $\vec y$ and $\vec s$ will be considered as fixed parameters, whereas $\zeta$ and $x$ will be variables. For the ease of notation, we omit the dependence on the parameters in our notations and write, for instance, $\Psi(\zeta;x)$ instead of $\Psi(\zeta;x,\vec y,\vec s)$.

\paragraph{Derivation of the Lax pair.}

We first show that the solution $\Psi(\zeta;x)$ of the RH problem solves systems of linear differential equations in $\zeta$ and $x$.
To that end, we first need to refine the asymptotic behavior of $\Psi(\zeta;x)$ for $\zeta$ near the singularities $0=y_k < ... < y_1$. Define, for $j = 1, ..., k$, $F_j$ by
\begin{equation}
\label{eq:psineary}
\Psi(\zeta;x) = F_j(\zeta;x) \left(I + \frac{s_{j+1} - s_j}{2\pi i} \sigma_+ \log(\zeta-y_j)\right)W_j(\zeta), 
\end{equation}
for $\zeta$ near $y_j$,
with $\sigma_+=\begin{pmatrix} 0 & 1 \\ 0 & 0 \end{pmatrix}$ and with $W_j$ given by
\begin{equation}
\label{eq:Vmatrix}
W_j(\zeta) = \left\{ \begin{array}{ll}
I, & \textrm{if } \zeta \in {\rm I}, \\
\begin{pmatrix} 1 & -s_j \\ 0 & 1 \end{pmatrix}, & \textrm{if } \zeta \in {\rm IV},
\end{array}\right.
\end{equation}
if $j < k$,
and
\begin{equation}
\label{eq:Wmatrix}
W_k(\zeta) = \left\{ \begin{array}{ll}
I, &  \textrm{if } \zeta \in {\rm I}, \\
\begin{pmatrix} 1 & 0 \\ -1 & 1 \end{pmatrix}, & \textrm{if } \zeta \in {\rm II}, \\
\begin{pmatrix} 1-s_k & -s_k \\ 1 & 1 \end{pmatrix}, & \textrm{if } \zeta \in {\rm III}, \\
\begin{pmatrix} 1 & -s_k \\ 0 & 1 \end{pmatrix}, & \textrm{if } \zeta \in {\rm IV},
\end{array}
\right.
\end{equation}
where I, II, III and IV are sectors delimited by the jump contours of $\Psi$, as illustrated in Figure \ref{fig:modelRHcontours}.
Using the jump relations for $\Psi$, it is straightforward to verify that $F_j$ is an analytic function of $\zeta$ near $y_j$. Indeed, the factors at the right of $F_j(\zeta;x)$ in \eqref{eq:psineary} are designed precisely to model the correct jumps for $\Psi$.

In order to simplify the Lax pair, we left-multiply $\Psi$ by a convenient uppertriangular matrix independent of $\zeta$ and define
\begin{equation}\label{eq:defPhi}
\Phi(\zeta;x) = e^{\frac{1}{4} \pi i \sigma_3}\begin{pmatrix} 1 & -ip(x) \\ 0 & 1 \end{pmatrix} \Psi(\zeta;x).
\end{equation}
Since $\Phi$ has the same jump matrices as $\Psi$, piecewise constant in $\zeta$ and independent of $x$, it follows that 
the matrices $A$ and $B$ defined by
\begin{equation}
\label{eq:Laxpair}
\left\{
\begin{matrix}
\Phi_\zeta(\zeta;x) = A(\zeta;x)\Phi(\zeta;x), \\
\Phi_x(\zeta;x) = B(\zeta;x)\Phi(\zeta;x),
\end{matrix}
\right.
\end{equation}
where the indices $\zeta$ and $x$ denote derivatives with respect to $\zeta$ and $x$, are meromorphic functions of $\zeta$ in the whole complex plane, with possible poles only at the points $y_1,\ldots, y_k$.

It follows from 
\eqref{eq:psineary} that $\Psi_x(\zeta;x) \Psi^{-1}(\zeta;x)$ and $\Phi_x(\zeta;x) \Phi^{-1}(\zeta;x)$ have removable singularities at $y_1, ..., y_k$, hence $B$ is an entire function of $\zeta$.
From the asymptotics for $\Psi$ at infinity, \eqref{eq:psiasympinf}, we deduce that $\Psi_x\Psi^{-1}$ takes the form
\[\Psi_x(\zeta;x) \Psi^{-1}(\zeta;x) = -i\zeta \sigma_+ + \begin{pmatrix} -p(x) & -2iq(x) \\ i & p(x) \end{pmatrix}.\]
By \eqref{eq:defPhi}, we get
\[B(\zeta;x) = \zeta \sigma_+ + \begin{pmatrix} 0 & 2q(x)+p^2(x)+p_x(x) \\ 1 & 0 \end{pmatrix}.\]
From the $1/\zeta$-term in the large $\zeta$ expansion for $B$ obtained using \eqref{eq:psiasympinf}, we get in addition the identity
\begin{equation}
\label{eq:psixpsiinversesubsubleadingtermidt}
p_x(x)-2q(x)-p^2(x) = 0.
\end{equation}
We may thus write $B$ as
\begin{equation}\label{eq:defB}B(\zeta;x) = \zeta \sigma_+ + \begin{pmatrix} 0 & 2p_x(x) \\ 1 & 0 \end{pmatrix}.\end{equation}

The form of the matrix $A(\zeta;x)$ is a bit more involved, as it follows from \eqref{eq:psineary} that $A$ has simple poles at the  points $y_1,\ldots, y_k$, because of the logarithmic singularities.
Using the asymptotics \eqref{eq:psiasympinf} of $\Psi$ at infinity, we get \[\Psi_\zeta(\zeta;x) \Psi^{-1}(\zeta;x) = -i\zeta\sigma_+ + i\begin{pmatrix} ip(x) & -2q(x)-\frac{x}{2} \\ 1 & -ip(x) \end{pmatrix} + \Or\left(\zeta^{-1}\right),\qquad \zeta\to\infty.\]
Therefore, using \eqref{eq:defPhi}, we have that the matrix $A$ has the form
\begin{equation}
A(\zeta;x) = \zeta \sigma_+ + \begin{pmatrix} 0 & 2q(x)+\frac{x}{2}+p^2(x) \\ 1 & 0 \end{pmatrix} + \sum_{j = 1}^k \frac{1}{\zeta-y_j} A_j(x)
\end{equation}
for matrices $A_1,\ldots, A_k$ independent of $\zeta$.
Using \eqref{eq:psixpsiinversesubsubleadingtermidt} again, we have
\begin{equation}
\label{eq:MatrixAGeneralFormula}
A(\zeta;x) = \zeta \sigma_+ + \begin{pmatrix} 0 & p_x(x)+\frac{x}{2} \\ 1 & 0 \end{pmatrix} + \sum_{j = 1}^k \frac{1}{\zeta-y_j} A_j(x).
\end{equation}

\paragraph{Compatibility condition.}
The Lax pair \eqref{eq:Laxpair} and the condition $\Phi_{x\zeta} = \Phi_{\zeta x}$ imply the compatibility condition \begin{equation}\label{eq:compatibility}B_\zeta - A_x + [B,A] = 0.\end{equation} This gives us several useful relations between $p$ and the matrices $A_j$. Since the determinant of $\Phi$ is identically equal to $1$, the trace of $A$ is zero, and we can write \begin{equation}\label{eq:Aabc}A(\zeta;x) = \begin{pmatrix} a(\zeta;x) & b(\zeta;x) \\ c(\zeta;x) & -a(\zeta;x) \end{pmatrix}.\end{equation} Using this parametrization, the compatibility condition reads
\begin{equation}
\label{eq:abcode}
\left\{
\begin{array}{l}
b = c(\zeta+2p_x) - \frac{1}{2} c_{xx} \\
a = \frac{1}{2} c_{x} \\
b_x = 1 - 2 a (\zeta+2p_x).
\end{array}
\right.
\end{equation}
We get from \eqref{eq:MatrixAGeneralFormula} that $c$ is of the form
\begin{equation}\label{eq:cexpansion}c(\zeta;x) = 1 + \sum_{j = 1}^k \frac{c_j(x)}{\zeta-y_j},\end{equation}
and that $b$ is of the form
\[b(\zeta;x) = \zeta + \frac{2p_x(x)+x}{2} + \sum_{j = 1}^k \frac{b_j(x)}{\zeta-y_j}.\]
We obtain from the first equation in \eqref{eq:abcode}, in particular from the constant term in the expansion as $\zeta\to\infty$, that
\begin{equation}
\label{eq:c0+c1}
\sum_{j = 0}^k c_j(x) = \frac{x}{2} - p_x(x).
\end{equation}

\bigskip

We now expand $(\zeta-y_j)^2\det \left(  A(\zeta;x) \right)$ for $\zeta$ near $y_j$, and we get from the constant term that
\[-\frac{1}{4} \frac{c_j'(x)^2}{c_j(x)^2} + \frac{1}{2} \frac{c_j''(x)}{c_j(x)}- y_j - 2p_x(x) = 0.\]
The fact that the right hand side is $0$ follows from the asymptotics of $\Psi$ as $\zeta\to y_j$, see \eqref{eq:psineary}. The relation \eqref{eq:c0+c1} yields the following system of equations for $c_1,..., c_k$,
\begin{equation}
\label{eq:c0c1equations}
\left\{
\begin{array}{l}
-\frac{1}{4} \frac{(c_1')^2}{c_1^2} + \frac{1}{2} \frac{c_1''}{c_1} + 2\sum_{j=1}^k c_j - x -y_1 = 0, \\
\vdots \\
-\frac{1}{4} \frac{(c_k')^2}{c_k^2} + \frac{1}{2} \frac{c_k''}{c_k} + 2\sum_{j=1}^k c_j - x -y_k = 0. \\
\end{array}
\right.
\end{equation}
We now make the transformation 
\begin{equation}\label{eq:def uc}
c_j(\xi+x_k;\vec y,\vec s) = -u_j^2(\xi;\vec x,\vec s),\qquad j=1,\ldots, k,\qquad y_j=x_j-x_k,\end{equation} and the system of equations for $c_1,...,c_k$ then transforms to the system \eqref{eq:theoremsystem} for $u_1,...,u_k$.

\paragraph{Expressions for $u_j$ in terms of $\Psi$.}


\bigskip

Next, we want to relate the functions $u_1,\ldots, u_k$ directly to the RH solution $\Psi$. By \eqref{eq:psineary} and \eqref{eq:defPhi}, we can expand $\Phi$ as $\zeta\to y_j$ in the following way,
\[\Phi(\zeta;x) = E_{0,j}(x)(I + E_{1,j}(x)(\zeta-y_j) + \Or((\zeta-y_j)^2)) \left( I + \frac{s_{j+1}-s_j}{2\pi i} \sigma_+ \log(\zeta-y_j) \right) W_j(\zeta),\]
with $E_{0,j}$ and $E_{1,j}$ depending on $x$ but not on $\zeta$. 

From the definition of $c_j(\xi+x_k) = -u_j^2(\xi)$, see \eqref{eq:Laxpair}, \eqref{eq:Aabc}, and \eqref{eq:cexpansion}, we also have the following expression for $u_j$ in terms of $\Phi$,
\begin{equation}
\label{eq:c0aslimit}
u_j^2(x-x_k) =-\lim_{\zeta \rightarrow y_j} (\zeta-y_j) \left(\left(\Phi_\zeta(\zeta;x)\right) \Phi^{-1}(\zeta;x)\right)_{2,1}.
\end{equation}

This allows us to express $u_j$ as
\begin{equation}\label{eq:idujxj}u_j^2(x-x_k) =  \frac{s_{j+1}-s_j}{2\pi i} (E_{0,j})_{2,1}^2(x)=-\frac{s_{j+1}-s_j}{2\pi } \Psi_{2,1}^2(y_j;x),\end{equation}
where we used \eqref{eq:defPhi}.

\medskip

On the other hand, exploiting the constant and  linear terms as $\zeta\to y_j$ in the equation $\Phi_x = B \Phi$, we find that
\[(E_{1,j}')_{2,1} = -(E_{0,j})_{2,1}^2,\]
where $'$ is the $x$-derivative.
It can be verified that \[(\Psi^{-1} \Psi_\zeta)_{2,1} (y_j;x) = (E_{1,j})_{2,1}(x),\] and we finally get the following relation, 
\begin{equation}
\label{eq:psi-c0_link}
\frac{\partial}{\partial x}(\Psi^{-1} \Psi_\zeta)_{2,1} (y_j;x) =-(E_{0,j})_{2,1}^2(x)= -\frac{2\pi i}{s_{j+1}-s_j}u_j^2(x-x_k).
\end{equation}
By \eqref{eq:idujxj}
and \eqref{eq:psi-c0_link}, Proposition \ref{prop Psi u} is proved.

\paragraph{Symmetry relation.}
One verifies easily that $\sigma_3\overline{\Psi(\overline\zeta;x)}\sigma_3$ satisfies the same RH conditions as $\Psi$. Since the solution to the RH problem for $\Psi$ is unique, we get the relation
\begin{equation}
\sigma_3\overline{\Psi(\overline\zeta;x)}\sigma_3=\Psi(\zeta;x).
\end{equation}
This implies that $\Psi_{2,1}(\zeta;x)$ is purely imaginary for real $\zeta$, hence $u_j^2(x)$ is always real by \eqref{eq:idujxj}. If $s_{j+1}>s_j$, we have that $u_j(x)$ is real; if $s_{j+1}<s_j$, $u_j(x)$ is purely imaginary.

\subsection{Proof of \eqref{eq:theoremformula}}

Define 
\begin{equation}
\widetilde F(x_k;\vec y,\vec s)=F(\vec x;\vec s),\qquad y_j=x_j-x_k.
\end{equation}
As $x_k\to \infty$ with $\vec y, \vec s$ fixed, 
$\widetilde F(x;\vec y,\vec s)$ converges to $1$, and is increasing. Therefore, by Proposition \ref{prop: diff id FPsi}, we have
\begin{multline}
\ln \widetilde F(\widetilde x_k;\vec y, \vec s)=-\int_{\widetilde x_k}^{+\infty}\frac{\partial}{\partial x_k}\ln\tilde F(x_k;\vec y,\vec s)dx_k=-\int_{\widetilde x_k}^{+\infty}\sum_{j=1}^k\frac{\partial}{\partial x_j}\ln F(\vec x,\vec s)d x_k\\=-\int_{\widetilde x_k}^{+\infty}\left(\sum_{j=1}^k\frac{s_{j+1}-s_j}{2\pi i}\left(\Psi^{-1}\Psi_\zeta\right)_{2,1}(\zeta=y_j;x,\vec y,\vec s) \right)dx.
\end{multline}
We now use Proposition \ref{prop Psi u} to integrate by parts, and obtain
\begin{multline}
\ln F(\vec x;\vec s)=\ln \widetilde F(x_k;\vec y,\vec s)=-\int_{x_k}^{+\infty}(\xi-x_k)\sum_{j=1}^k u_j^2(\xi-x_k;\vec x,\vec s) d\xi\\=-\int_{0}^{+\infty}\xi\sum_{j=1}^k u_j^2(\xi;\vec x,\vec s) d\xi,
\end{multline}
and we get formula \eqref{eq:theoremformula}. To complete the proof of Theorem \ref{thm:fredholmdetrep}, it only remains to  prove the asymptotics \eqref{eq:asujAiry}.

\section{Asymptotic analysis of the RH problem for $\Psi$}\label{section:RH}

In this section, we study the RH problem for $\Psi$, which depends on $x$, $\vec y$, $\vec s$, asymptotically.
There are three asymptotic regimes that we will study:
\begin{enumerate}
\item $x\to +\infty$ while keeping $\vec y$, $\vec s$ fixed,
\item $x_j-x_{j+1}\to 0$ with all other parameters fixed, for $j=0,\ldots k-1$, where we set $x_0=+\infty$,
\item $s_j\to s_{j+1}$ with all other parameters fixed, for $j=1,\ldots, k$, with $s_{k+1}=1$.
\end{enumerate}
Each of these cases will require a separate analysis of the RH problem for $\Psi$, using Deift-Zhou steepest descent techniques \cite{DeiftZhou}. The RH analysis in the first case $x\to +\infty$ requires several transformations and the construction of parametrices; the other cases are more straightforward. The asymptotic analysis of $\Psi$ will enable us to understand the asymptotic behavior of the solutions $u_j$ to the system \eqref{eq:theoremsystem}, and to prove \eqref{eq:asujAiry} and Theorem \ref{thm: asymptotics}.

\subsection{Asymptotic analysis as $x\to +\infty$}\label{sec:Psi-xtoinfty}

We will perform a series of invertible transformations $\Psi \mapsto A \mapsto B \mapsto C \mapsto D$ in order to obtain a matrix $D$ close to the identity matrix.

\subsubsection{Re-scaling and shift of the contour}

The first transformation $\Psi \mapsto A$ consists of a re-scaling of the $\zeta$-variable: we define
\begin{equation}\label{eq:defA}A(\lambda;x,\vec y,\vec s) = x^{-\frac{\sigma_3}{4}} \Psi(x\lambda;x,\vec y,\vec s).\end{equation}

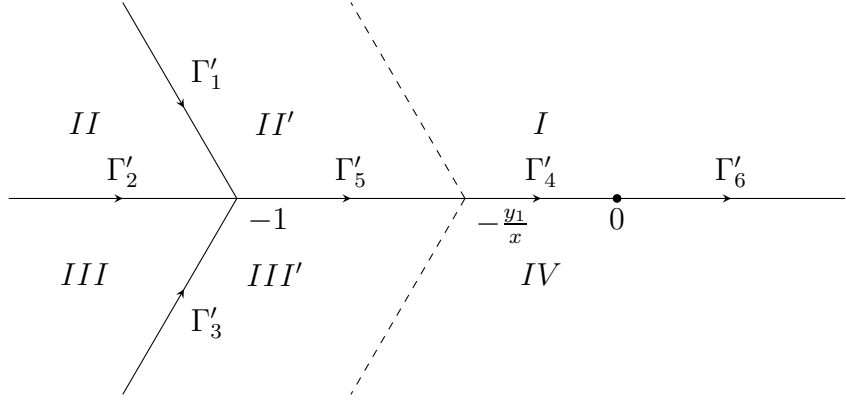
\begin{figure}
\centering
\begin{tikzpicture}[> = stealth]
\draw [->- = .5] (-6,0) -- node [anchor = south] {$\Gamma_2'$} (-3,0);
\draw [->- = .5] (-3,0) -- node [anchor = south] {$\Gamma_5'$} (0,0);
\draw [->- = .5] (0,0) -- node [anchor = south] {$\Gamma_4'$} (2,0);
\draw [->- = .5] (2,0) -- node[anchor = south] {$\Gamma_6'$} (5,0);
\draw [-<- = .5] (-3,0) -- node [anchor = south west] {$\Gamma_1'$} ++(120:3cm);
\draw [-<- = .5] (-3,0) -- node [anchor = north west] {$\Gamma_3'$} ++(240:3cm);
\draw [dashed] (0,0) -- (120:3cm);
\draw [dashed] (0,0) -- (240:3cm);

\draw (-5,1) node {$II$};
\draw (-2.5,1) node {$II'$};
\draw (1,1) node {$I$};
\draw (-5,-1) node {$III$};
\draw (-2.5,-1) node {$III'$};
\draw (1,-1) node {$IV$};

\draw (-3,0) node [anchor = north west] {$-1$};
\draw (0,0) node [anchor = north west] {$-\frac{y_1}{x}$};
\draw (2,0) node [anchor = north] {$0$};
\filldraw (2,0) circle (0.05cm);
\end{tikzpicture}
\caption{The shifted jump contour $\Gamma'$ for $C$ (solid lines) compared to the contour $\Gamma$ for $B$ (dashed lines).}
\label{fig:BandCjumpcontours}
\end{figure}

The second transformation $A \mapsto B$ shifts the two non-real parts of the jump contour to the left, in order to separate them from the discontinuities in the jump matrices which are now located at $\lambda_j:=y_j/x$, $j=1,\ldots, k$, with $\lambda_k=y_k=0$. To do this, we need to continue $A$ analytically from sectors I, IV to regions II', III', see Figure \ref{fig:BandCjumpcontours}.

More precisely, we define the matrix $B$ by
\begin{equation}\label{eq:defB3}
B(\lambda)=B(\lambda;x,\vec y,\vec s) := A(\lambda;x,\vec y,\vec s) \left\{ \begin{array}{l}
\begin{pmatrix} 1 & 0 \\ 1 & 1 \end{pmatrix}, \textrm{ for } \lambda \in {\rm II'}, \\
\begin{pmatrix} 1 & 0 \\ -1 & 1 \end{pmatrix}, \textrm{ for } \lambda \in {\rm III'}, \textrm{ and} \\
\begin{pmatrix} 1 & 0 \\ 0 & 1 \end{pmatrix}, \textrm{ everywhere else.}
\end{array}
\right.
\end{equation}
The function $B$ satisfies the following RH problem for $B$. 

\paragraph{RH problem for $B$.}

\begin{itemize}
\item[(a)] $B : \C \backslash \Gamma' \rightarrow \C^{2\times 2}$ is analytic, with $\Gamma'=\mathbb R\cup \left(-1+e^{\pm\frac{2\pi i}{3}}\mathbb R^+\right)$,
\item[(b)] $B$ has continuous boundary values on $\Gamma' \backslash \{-1,\lambda_1,\ldots, \lambda_k\}$ with $\lambda_j=y_j/x$, and $B_+(\lambda) = B_-(\lambda) J_B(\lambda)$, where $J_B$ is given by
\[J_B(\lambda) = \left\{ \begin{array}{ll}
\begin{pmatrix} 1 & 0 \\ 1 & 1 \end{pmatrix} &\textrm{ on } -1+e^{\pm\frac{2\pi i}{3}}\mathbb R^+, \\
\begin{pmatrix} 0 & 1 \\ -1 & 0 \end{pmatrix} &\textrm{ on } (-\infty,-1), \\
\begin{pmatrix} 1 & s_j \\ 0 & 1 \end{pmatrix} &\textrm{ on } (\lambda_j, \lambda_{j-1}),\ j=1,\ldots, k,\ \lambda_0=+\infty, \\
\begin{pmatrix} 1 & 1 \\ 0 & 1 \end{pmatrix} &\textrm{ on } (-1,\lambda_k),\end{array}
\right.
\]
\item[(c)] $B$ has the following asymptotics:
\[B(\lambda) = \left( I + \Or\left( \lambda^{-1} \right) \right) \lambda^{\sigma_3/4} M^{-1} e^{-x^{3/2} \left( \frac{2}{3} \lambda^{3/2} + \lambda^{1/2} \right)}, \textrm{ as } \lambda \rightarrow \infty, \]
\item[(d)] $B$ has the following behavior at $\lambda_j$ for $j=1,\ldots, k$:
\[B(\lambda) = \widehat F_j(\lambda) \left( I + \frac{s_{j+1}-s_j}{2\pi i} \sigma_+ \log\left(\lambda -\lambda_j \right) \right) \widehat W_j(\lambda), \textrm{ as } \lambda \to \lambda_j, \]
where $\widehat F_j$ is analytic in a neighborhood of $\lambda_j$, $\widehat W_j = I$ for $\lambda$ in the upper half plane, and $\widehat W_j= \begin{pmatrix} 1 & -s_j \\ 0 & 1 \end{pmatrix}$ for $\lambda$ in the lower half plane, similarly to \eqref{eq:psineary}.
\end{itemize}


\subsubsection{Normalization at infinity}

The goal of the next transformation is to normalize the RH problem at infinity, in such a way that the jump matrices behave nicely. To that end, we define
\begin{equation}\label{eq:defg}
g(\lambda)= \frac{2}{3} (\lambda+1)^{3/2}, \end{equation}
with branch cut on $(-\infty,-1]$, such that $g_-(\lambda) = -g_+(\lambda)$ on $(-\infty, -1)$.
At infinity, $g$ behaves as
\begin{equation}\label{eq:asg} g(\lambda) = \frac{2}{3} \lambda^{3/2} + \lambda^{1/2} + \frac{1}{4} \lambda^{-1/2} + \Or(\lambda^{-3/2} ), \textrm{ as } \lambda \to \infty, \end{equation}
with all branch cuts at the right hand side on $(-\infty,0]$.
We now perform the transformation $B \mapsto C$ by defining
\begin{equation}\label{eq:defC} C(\lambda) = \begin{pmatrix} 1 & -\frac{i x^{3/2}}{4} \\ 0 & 1 \end{pmatrix} B(\lambda) e^{x^{3/2} g(\lambda) \sigma_3}. \end{equation}
The constant upper-triangular pre-factor in the definition of $C$ is needed to ensure that $C$ has a suitable behavior at infinity. $C$ satisfies the following RH problem.

\paragraph{RH problem for $C$.}

\begin{itemize}
\item[(a)] $C : \C \backslash \Gamma' \rightarrow \C^{2\times 2}$ is analytic,
\item[(b)] $C_+=C_-J_C$ on $\Gamma'\setminus\{-1,\lambda_1,\ldots, \lambda_k\}$ with the jump matrices $J_C$ given by \begin{equation}J_C(\lambda)=e^{-x^{3/2} g_-(\lambda) \sigma_3} J_B(z) e^{x^{3/2} g_+(\lambda) \sigma_3},\end{equation} i.e.
\[ J_C(\lambda) = \left\{ \begin{array}{ll}
\begin{pmatrix} 1 & 0 \\ e^{2x^{3/2} g(\lambda)} & 1 \end{pmatrix} &\textrm{ on } -1+e^{\pm\frac{2\pi i}{3}}\mathbb R^+, \\
\begin{pmatrix} 0 & 1 \\ -1 & 0 \end{pmatrix} &\textrm{ on } (-\infty,-1), \\
\begin{pmatrix} 1 & e^{-2x^{3/2} g(\lambda)} \\ 0 & 1 \end{pmatrix} &\textrm{ on } (-1,\lambda_k),\\
\begin{pmatrix} 1 & s_j e^{-2x^{3/2} g(\lambda)} \\ 0 & 1 \end{pmatrix} &\textrm{ on } (\lambda_{j},\lambda_{j-1}),\ j=1,\ldots, k,
\end{array}\right. \]
\item[(c)] $C$ has the following asymptotics at $\infty$:
\[C(\lambda) = \left( I + \Or\left( \lambda^{-1} \right) \right) \lambda^{\sigma_3/4} M^{-1}, \textrm{ as } \lambda \to \infty, \]
\item[(d)] $C$ has the same logarithmic behavior near $\lambda_j$ as $B$ has, multiplied to the right by $e^{x^{3/2} g(\lambda) \sigma_3}$.
\end{itemize}


We now need to construct a global parametrix and local parametrices near the special points $-1$ and $0$.

\subsubsection{Global parametrix}

Away from $-1, 0$, all jumps of $C$ except the one on $(-\infty, -1)$ tend to the identity matrix as $x \to +\infty$. Ignoring small jumps and small neighbourhoods of $-1, 0$, we get the following.
\paragraph{RH problem for $C^{(\infty)}$.}
\begin{itemize}
\item[(a)] $C^{(\infty)}$ is analytic everywhere in the complex plane except on $(-\infty, -1]$,
\item[(b)] $C^{(\infty)}_+=C^{(\infty)}_-\begin{pmatrix} 0 & 1 \\ -1 & 0 \end{pmatrix}$ on $(-\infty,-1)$,
\item[(c)] $C^{(\infty)}$ has the following asymptotics:
\[ C^{(\infty)}(\lambda) = \left( I + \Or\left(\lambda^{-1}\right) \right) \lambda^{\sigma_3/4} M^{-1}, \textrm{ as } \lambda \to \infty. \]
\end{itemize}

A simple solution to this problem is given by
\begin{equation}C^{(\infty)}(\lambda) = (\lambda+1)^{\sigma_3/4} M^{-1},
\label{eq:defCinf}\end{equation} 
with the branch cut of $(\lambda+1)^{\sigma_3/4}$ along $(-\infty,-1]$.

\subsubsection{Local parametrix near $-1$}

Near $-1$, the RH problem for $C$ resembles the standard model RH problem built out of the Airy function, and we will use this model problem to build the local parametrix. This construction is fairly standard, and was also used in the proof of Proposition \ref{prop: diff id FPsi}, see \eqref{def:PhiA}--\eqref{eq:asPhi}.
We will construct the local parametrix in the form
\begin{equation}\label{eq:defC-1}C^{(-1)}(\lambda) = E(\lambda) \widehat\Phi(x^{3/2} (\lambda+1)) e^{x^{3/2} g(\lambda) \sigma_3},\end{equation}
where $E$ is analytic around $-1$ and is chosen below, and $\widehat\Phi$ is, analogously to $\Phi$ defined in \eqref{def:Phi} but with a slightly different jump contour, the solution to the standard Airy model RH problem, given by
\begin{equation}\label{def:hatPhi}
\widehat\Phi(z)=\begin{cases}\Phi_{A}(z),& \mbox{ for }z \mbox{ in region I,}\\
\Phi_{B}(z),& \mbox{ for }z \mbox{ in region II,}\\
\Phi_{C}(z),& \mbox{ for }z \mbox{ in region III,}\\
\Phi_{D}(z),& \mbox{ for }z \mbox{ in region IV,}
\end{cases}
\end{equation}
where the regions I, II, III and IV are the ones described in Figure \ref{fig:modelRHcontours}. The asymptotics for $\widehat\Phi$ as $z\to\infty$ are the same as the ones for $\Phi$ given in \eqref{eq:asPhi}.

\bigskip

Let $0 < \delta < 1/2$, $U_{-1,\delta}$ the disk of radius $\delta$ around $-1$, and define $C^{(-1)} : U_{-1,\delta} \backslash \Gamma' \rightarrow \C^{2\times 2}$ by
\[C^{(-1)}(\lambda) = e^{\frac{i\pi}{4} \sigma_3} \begin{pmatrix} 0 & -1 \\ 1 & 0 \end{pmatrix} (x^{3/2})^{\sigma_3/4} \widehat\Phi(x(\lambda+1)) e^{x^{3/2} g(\lambda) \sigma_3}.\]
On the boundary of $U_{-1,\delta}$, we have 
\begin{equation}\label{eq:matching}C^{(-1)}(\lambda) = \left( I + \Or\left( x^{-1} \right) \right) C^{(\infty)}(\lambda),
\end{equation} uniformly for $\lambda \in \partial U_{-1,\delta}$ as $x \rightarrow + \infty$. Moreover, $C^{(-1)}$ has the same jumps as $C$ on $\Gamma' \cap U_{-1,\delta}$.

\subsubsection{Local parametrix near 0}

When $x$ gets large and the values $y_j$ remain fixed, then the discontinuities $\lambda_j$ are all close to $0$. We can construct a local parametrix near $0$ which models the jumps near $0$ in an elementary way. Consider the function $v$ which takes the value $s_j$ on $(\lambda_j, \lambda_{j-1})$ for $j = 1,..., k+1$, with the convention that $\lambda_0 = +\infty$, $\lambda_{k+1} = -1/2$ and $s_{k+1} = 1$. Next we define the matrix-valued function $C^{({\rm loc})}$ by
\begin{equation}C^{({\rm loc})}(\lambda) = \begin{pmatrix} 1 & \int_{-\frac{1}{2}}^{+\infty} \frac{v(w)}{w-\lambda} e^{-2x^{3/2}g(w)} dw \\ 0 & 1 \end{pmatrix}.\label{eq:defCloc}\end{equation}This function has the same jumps as $C$ across the real axis, but does not have the correct matching with the parametrix at infinity $C^{(\infty)}$ on the boundary of $U_{0,\delta}$. The matrix $C^{({\rm loc})}$ is however exponentially close to the identity matrix on this boundary, and we may therefore construct the local parametrix near $0$ by defining
\begin{equation}\label{eq:defC0} C^{(0)}(\lambda) = C^{(\infty)}(\lambda)C^{({\rm loc})}(\lambda),\quad
\lambda \in U_{0,\delta}.\end{equation}
In this way, since $C^{(\infty)}$ is analytic near $0$, $C^{(0)}$ has the same jumps as $S$ has near $0$ and it matches with $C^{(\infty)}$ on the boundary of $U_{0,\delta}$,
\begin{equation}
\label{eq:matching0}
C^{(-1)}(\lambda) = \left( I + \Or\left( x^{-1} \right) \right) C^{(\infty)}(\lambda),\qquad x\to +\infty,
\end{equation}
uniformly for $\lambda$ on the boundary of $U_{0,\delta}$.

\subsubsection{Last transformation $C \mapsto D$}

\begin{figure}
\centering
\begin{tikzpicture}[> = stealth]
\draw[->- = .5] (120:3cm) -- node [anchor = south west] {$\Gamma_1^D$} (120:.5cm);
\draw[->- = .5] (240:3cm) -- node [anchor = north west] {$\Gamma_3^D$} (240:.5cm);

\draw (0,0) node {$-1$};

\draw[->- = .5] (.5,0) arc (0:360:.5cm);
\draw (-.5,0) node [anchor = east] {$\Gamma_2^D$};

\draw (4,0) node {$0$};

\draw[->- = .5] (.5,0) -- node[anchor = south] {$\Gamma_5^D$} (3.5,0);
\draw[->- = .25] (4.5,0) arc (0:360:.5cm);
\draw (4,.5) node [anchor = south] {$\Gamma_6^D$};

\draw[->- = .5] (4.5,0) -- node [anchor = south] {$\Gamma_4^D$} (8,0);
\end{tikzpicture}
\caption{Jump contours for $D$}
\label{fig:jumpcontoursD}
\end{figure}
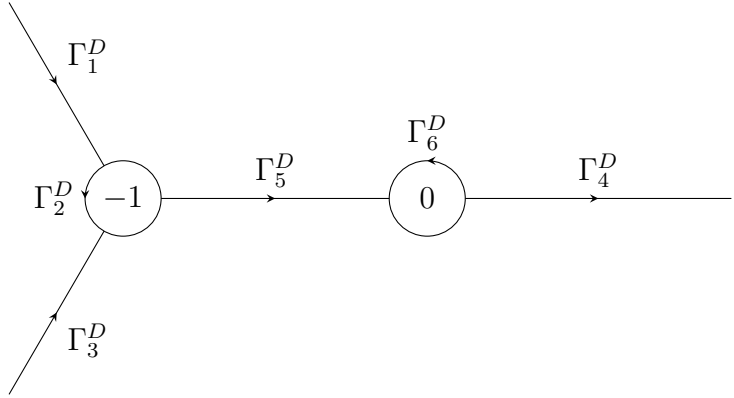

We take $0 < \delta < 1/2$ and $x$ large enough, such that $\lambda_j < \delta$ for all $j$. We define the matrix-valued function $D$ by
\begin{equation}\label{eq:defD}D(\lambda) = \left\{ \begin{array}{ll}
C(\lambda) C^{(-1)}(\lambda)^{-1} & {\rm for}\ \lambda \in U_{-1,\delta}, \\
C(\lambda) C^{(0)}(\lambda)^{-1} & {\rm for}\ \lambda \in U_{0,\delta}, \\
C(\lambda) C^{(\infty)}(\lambda)^{-1} & {\rm otherwise.}
\end{array}\right.\end{equation}
The matrix $D$ satisfies the following RH problem. The contours $\Gamma_1^D, ..., \Gamma_6^D$ are depicted on Figure \ref{fig:jumpcontoursD}.

\paragraph{RH problem for $D$.}

\begin{itemize}
\item[(a)] $D : \C \backslash \cup_{j=1}^6 \Gamma_j^D \longrightarrow \C^{2\times 2}$ is analytic, with $\Gamma_j^D$, $j=1,\ldots, 6$, as in Figure \ref{fig:jumpcontoursD},
\item[(b)] $D$ has the jumps $D_+=D_-J_D$ on $\cup_{j=1}^6 \Gamma_j^D$, with
\begin{align}
&J_D(\lambda)=\begin{pmatrix} 1 & 0 \\ e^{2x^{3/2} g(\lambda)} & 1 \end{pmatrix}&&\mbox{on $\Gamma_1^D$ and $\Gamma_3^D$,}\\
&J_D(\lambda)= C^{(\infty)}(\lambda) C^{(-1)}(\lambda)^{-1}&&\mbox{on $\Gamma_2^D$,}\\ &J_D(\lambda)=\begin{pmatrix} 1 & s_k e^{-2x^{3/2} g(\lambda)} \\ 0 & 1 \end{pmatrix}&&\mbox{on $\Gamma_4^D$,}\\
&J_D(\lambda)=\begin{pmatrix} 1 & e^{-2x^{3/2} g(\lambda)} \\ 0 & 1 \end{pmatrix}
&&\mbox{on $\Gamma_5^D$,}\\
&\label{eq:jump D0}J_D(\lambda)=C^{(\infty)}(\lambda) C^{({\rm loc})}(\lambda)^{-1} C^{(\infty)}(\lambda)^{-1}&&\mbox{on $\Gamma_6^D$,}
\end{align}
\item[(c)]$D$ has the following asymptotic behavior at infinity:
\[D(\lambda) = I + \Or\left( \lambda^{-1} \right), \qquad {\rm as}\, \lambda \rightarrow \infty.\]
\end{itemize}

As $x(\lambda_k+1) \to +\infty$, all the jumps of $D$ are $I + \Or(x^{-1}(\lambda_k+1)^{-1})$, uniformly for $\lambda$ on the jump contours. This implies that the function $D$ satisfies
\begin{equation}D(\lambda) = I + \Or\left( x^{-1}\right),\qquad \textrm{as } x \to +\infty,\label{eq:as D}\end{equation}
uniformly for $\lambda\in\C \backslash \cup_{j=1}^6 \Gamma_j^D$.

\subsubsection{Asymptotic behavior of $u_0,...,u_k$.}

By inverting the transformations $\Psi\mapsto A\mapsto B\mapsto C\mapsto D$, the large $x$ asymptotics \eqref{eq:as D} for $D$ enable us to find large $x$ asymptotics for $\Psi$ in terms of the parametrices $C^{(\infty)}$, $C^{(-1)}$, and $C^{(0)}$.
Moreover, using the identity \eqref{eq:uintermsofPsi21}, we can express the asymptotics of $u_j$, $j=1,\ldots, k$ in terms of the parametrices.

Using \eqref{eq:defA} and \eqref{eq:defB3}, we have
\begin{multline}
u_j^2(x-x_k;\vec x,\vec s)=
-\frac{s_{j+1}-s_j}{2\pi}\lim_{\zeta\to y_j}\Psi_{2,1}^2(\zeta;x,\vec y,\vec s)=-\frac{s_{j+1}-s_j}{2\pi\sqrt{x}}\lim_{\lambda\to \lambda_j}A_{2,1}^2(\lambda;x,\vec y,\vec s)\\=-\frac{s_{j+1}-s_j}{2\pi\sqrt{x}}\lim_{\lambda\to \lambda_j}B_{2,1}^2(\lambda;x,\vec y,\vec s),
\end{multline}
where we adopt the convention that the limits $\lambda\to\lambda_j$ are taken from region I, see Figure \ref{fig:BandCjumpcontours}. 
By \eqref{eq:defg}, \eqref{eq:defC}, and \eqref{eq:defD},
\begin{multline}
u_j^2(x-x_k;\vec x,\vec s)=-\frac{s_{j+1}-s_j}{2\pi\sqrt{x}}\lim_{\lambda\to \lambda_j}C_{2,1}^2(\lambda;x,\vec y,\vec s)e^{-\frac{4}{3}(y_j+x)^{3/2}}\\
=-\frac{s_{j+1}-s_j}{2\pi\sqrt{x}}\lim_{\lambda\to \lambda_j}\left(DC^{(0)}\right)_{2,1}^2(\lambda;x,\vec y,\vec s)e^{-\frac{4}{3}(y_j+x)^{3/2}}.
\end{multline}
Now we use the asymptotics \eqref{eq:as D} and the expression \eqref{eq:defC0} (see also \eqref{eq:defCinf} and \eqref{eq:defCloc}) for the local parametrix to obtain
\begin{multline}
u_j^2(x-x_k;\vec x,\vec s)=-\frac{s_{j+1}-s_j}{2\pi\sqrt{x}}\left(D_{2,1}(\lambda_j)C_{1,1, +}^{(0)}(\lambda_j)+D_{2,2}(\lambda_j)C_{2,1, +}^{(0)}(\lambda_j)\right)e^{-\frac{4}{3}(y_j+x)^{3/2}}\\
=\frac{s_{j+1}-s_j}{4\pi\sqrt{y_j+x}}e^{-\frac{4}{3}(y_j+x)^{3/2}}\left(1+\Or(x^{-1})\right),\qquad x\to +\infty.
\end{multline}
Using the fact that $y_j=x_j-x_k$ and the standard large argument asymptotics for the Airy function, we obtain \eqref{eq:asujAiry} as $x\to +\infty$.

\subsection{Asymptotic analysis  as $x_{j}\to x_{j-1}$}

Fix $j\in\{2,\ldots, k\}$, and denote $\vec y^{[j]}, \vec s^{[j]}$ for the vectors $\vec y, \vec s$ without their $j$-th components.
We also define 
\begin{equation}\Psi^{[j]}(\zeta;x,\vec y,\vec s)=\Psi(\zeta;x,\vec y^{[j]}, \vec s^{[j]}),
\end{equation}
i.e., $\Psi^{[j]}$ is the solution to the model RH problem for $\Psi$ with $k-1$ singularities located at $y_1, \ldots, y_{j-1}, y_{j+1},\ldots, y_k=0$.
The RH conditions for $\Psi$ and $\Psi^{[j]}$ are the same, except on $[x_j, x_{j-1}]$. Therefore, it is not surprising that if $x_j$ and $x_{j-1}$ are close to each other, the function $\Psi^{[j]}(\zeta;x,\vec y,\vec s)$ will be a good approximation for $\Psi(\zeta;x,\vec y,\vec s)$ as $x_j\to x_{j-1}$ for $\zeta$ not too close to $x_j, x_{j-1}$.

Near $x_j, x_{j-1}$, we need to build a local parametrix, for which we need to define (similarly as in \eqref{eq:psineary}) the function $F_{j-1}^{[j]}(\zeta)$ by the equation
\begin{equation}
\label{eq:psistarnearsingularities}
\Psi^{[j]}(\zeta;x,\vec y,\vec s) = F_{j-1}^{[j]}(\zeta) \left(I + \frac{s_{j+1} -  s_{j-1}}{2\pi i} \sigma_+ \log(\zeta-y_{j-1})\right)W_{j-1}(\zeta), 
\end{equation}
for $\zeta$ near $y_{j-1}$, with $W_{j-1}$ given as in \eqref{eq:Vmatrix}--\eqref{eq:Wmatrix}, and with principal branches of the logarithms. Then, one verifies, in a similar way as we did for \eqref{eq:psineary}, that $F_{j-1}^{[j]}$ is analytic at $y_{j-1}$.
We now define the local parametrix $P$ in the form
\begin{equation}\label{eq:defP}
P(\zeta;x,\vec y,\vec s) = F_{j-1}^{[j]}(\zeta) \left(I + \frac{s_{j+1} - s_j}{2\pi i} \sigma_+ \log(\zeta-y_{j})+\frac{s_{j} - s_{j-1}}{2\pi i} \sigma_+ \log(\zeta-y_{j-1})\right)W_{j-1}(\zeta),
\end{equation}
for $\zeta$ in a fixed sufficiently small open set $U$ which contains $y_j, y_{j-1}$ but which does not contain any of the other singularities $y_\ell$. It is then straightforward to check that $P$ has exactly the same jump relations as $\Psi$ has inside $U$, and
by \eqref{eq:psistarnearsingularities} and \eqref{eq:defP}, we  get the matching condition
\begin{equation}\label{eq:matchingPsix}
P(\zeta;x,\vec y,\vec s)\Psi^{[j]}(\zeta;x,\vec y,\vec s)^{-1}=I+\Or\left(x_{j-1}-x_{j}\right),\qquad x_{j}\to x_{j-1},
\end{equation}
uniformly for $\zeta\in \partial U$.

Define
\begin{equation}\label{eq:defR}
R(\zeta)=\begin{cases}
\Psi(\zeta;x,\vec x,\vec s)\Psi^{[j]}(\zeta;x,\vec y,\vec s)^{-1},&\zeta\in \mathbb C\setminus U,\\
\Psi(\zeta;x,\vec x,\vec s)P(\zeta;x,\vec y,\vec s)^{-1},&\zeta\in U.
\end{cases}
\end{equation}
Then $R$ is analytic everywhere in the complex plane, except on $\partial U$, where we have $R_+=R_-\left(I+\Or\left(x_{j-1}-x_{j}\right)\right)$ as $x_{j}\to x_{j-1}$ by \eqref{eq:matchingPsix}. As $\zeta\to\infty$, $R(\zeta)\to I$ because $\Psi$ and $\Psi^{[j]}$ have the same asymptotics \eqref{eq:psiasympinf}. It follows from usual small-norm arguments for RH problems that
\begin{equation}
R(\zeta)=I+\Or(x_{j-1}- x_{j}),\qquad x_{j}\to x_{j-1},
\end{equation}
uniformly for $\zeta\in\mathbb \C \setminus\partial U$.
Hence,  as $x_j\to x_{j-1}$, we have for $\ell<j-1$,
\begin{align*}
u_\ell^2(x-x_k;\vec x,\vec s)&=
-\frac{s_{\ell+1}-s_\ell}{2\pi}\lim_{\zeta\to y_\ell}\Psi_{2,1}(\zeta;x,\vec y,\vec s)^2\\&=-\frac{s_{\ell+1}-s_\ell}{2\pi}\lim_{\zeta\to \zeta_\ell}\Psi_{2,1}^{[j]}(\zeta;x,\vec y,\vec s)^2+\Or(x_{j-1}-x_j)\\&=u_\ell^2(x-x_k;\vec x^{[j]},\vec s^{[j]})+\Or(x_{j-1}-x_j),
\end{align*}
and similarly for $\ell>j$,
\[u_\ell^2(x-x_k;\vec x,\vec s)=u_{\ell-1}^2(x-x_k;\vec x^{[j]},\vec s^{[j]})+\Or(x_{j-1}-x_j).\]
For $\ell=j, j-1$, we have
\begin{align*}
u_\ell^2(x-x_k;\vec x,\vec s)&=
-\frac{s_{\ell+1}-s_\ell}{2\pi}\lim_{\zeta\to y_\ell}\Psi_{2,1}(\zeta;x,\vec y,\vec s)^2\\
&=-\frac{s_{\ell+1}-s_\ell}{2\pi}\lim_{\zeta\to y_\ell}P_{2,1}(\zeta;x,\vec y,\vec s)^2+\Or(x_{j-1}-x_j)\\
&=-\frac{s_{\ell+1}-s_\ell}{2\pi}F_{j-1}^{[j]}(y_\ell)_{2,1}^2+\Or(x_{j-1}-x_j),
\end{align*}
which implies that
\begin{multline*}
u_{j-1}^2(x-x_k;\vec x,\vec s)+u_j^2(x-x_k;\vec x,\vec s)=-\frac{s_{j+1}-s_{j-1}}{2\pi}F_{j-1}^{[j]}(y_j)_{2,1}^2+\Or(x_{j-1}-x_j)\\=u_j^2(x-x_k;\vec x^{[j]},\vec s^{[j]})+\Or(x_{j-1}-x_j)
\end{multline*}
as $x_{j}\to x_{j-1}$. This proves Theorem \ref{thm: asymptotics}, part 2.

Part 3 of Theorem \ref{thm: asymptotics} can be seen as a special case of the previous with $j=1$, if one identifies $x_0$ with $+\infty$. Then, $U$ has to be chosen as a neighborhood of $[y_1,+\infty)$, and the parametrix $P$ in $U$ has to be defined in a slightly different manner as 
\begin{equation}\label{eq:defP2}
P(\zeta;x,\vec y,\vec s) = 
\Psi^{[1]}(\zeta;x,\vec y,\vec s)W_{2}(\zeta)^{-1} \left(I + \frac{s_{2} - s_1}{2\pi i} \sigma_+ \log(\zeta-y_{1})\right)W_{2}(\zeta),
\end{equation}
with $W_2$ as in \eqref{eq:Vmatrix}--\eqref{eq:Wmatrix}, and with the global parametrix still given by $\Psi^{[1]}$.
$R$ is defined in the same way as before, and has a jump $I+\Or(x_1^{-1})$  on the boundary of $U$ as $x_1\to +\infty$. A straightforward calculation similar to the one for $j>1$ now leads to the proof of part 3 of Theorem \ref{thm: asymptotics}.

\subsection{Asymptotic analysis of the model RH problem for $\Psi$ as $s_{j+1}- s_{j}\to 0$}

Fix $j\in\{1,\ldots, k\}$, and define as before
\begin{equation}\Psi^{[j]}(\zeta;x,\vec y,\vec s)=\Psi(\zeta;x,\vec y^{[j]}, \vec s^{[j]}),
\end{equation}
with $\vec y^{[j]}, \vec s^{[j]}$ the vectors $\vec y, \vec s$ without their $j$-th components.
This function will again serve as the global parametrix for $\Psi$ as $s_{j+1}- s_{j}\to 0$.
The local parametrix in an open set $U$ containing both $y_j$ and $y_{j-1}$, but no other singularities,
is given by \eqref{eq:defP} if $j\neq 1$, and by \eqref{eq:defP2} if $j=1$, exactly as in the analysis for $x_j\to x_{j-1}$.

We define $R$ as in \eqref{eq:defR}, and 
$R$ is analytic everywhere in the complex plane, except on $\partial U$, where we have $R_+=R_-\left(I+\Or\left(s_{j}-s_{j+1}\right)\right)$ as $s_{j+1}- s_{j}\to 0$ by \eqref{eq:defP}. As $\zeta\to\infty$, $R(\zeta)\to I$ because $\Psi$ and $\Psi^{[j]}$ have the same asymptotics \eqref{eq:psiasympinf}. It follows that
\begin{equation}
R(\zeta)=I+\Or(s_{j+1}- s_{j}),\qquad s_{j+1}- s_{j}\to 0.
\end{equation}

It follows that, for $\ell< j$,
\begin{align*}
u_\ell^2(x-x_k;\vec x,\vec s)&=
-\frac{s_{\ell+1}-s_\ell}{2\pi}\lim_{\zeta\to y_\ell}\Psi_{2,1}(\zeta;x,\vec y,\vec s)^2\\
&=-\frac{s_{\ell+1}-s_\ell}{2\pi}\lim_{\zeta\to y_\ell}\Psi_{2,1}^{[j]}(\zeta;x,\vec y,\vec s)^2+\Or(s_{j+1}-s_j)\\
&=u_\ell^2(x-x_k;\vec x^{[j]},\vec s^{[j]})+\Or(s_{j+1}-s_j),
\end{align*}
as $s_{j+1}- s_{j}\to 0$, and a similar calculation applies to the case $\ell>j$.
For $\ell=j$, we have
\[u_j^2(x-x_k;\vec x,\vec s)=
-\frac{s_{j+1}-s_j}{2\pi}\lim_{\zeta\to y_j}\Psi_{2,1}(\zeta;x,\vec y,\vec s)^2=\Or(s_{j+1}-s_j),\]
as $s_{j+1}- s_{j}\to 0$. This completes the proof of part 1 of Theorem \ref{thm: asymptotics}.

\section*{Acknowledgements}
The authors are grateful to Max Atkin, Jinho Baik, and Andrew Hone for useful comments and discussions.
They were supported by the European Research Council under the European Union's Seventh Framework Programme (FP/2007/2013)/ ERC Grant Agreement 307074 and by the Belgian Interuniversity Attraction Pole P07/18.


\begin{thebibliography}{99}
\addcontentsline{toc}{section}{References}




\bibitem{AblowitzSegur} M.J. Ablowitz and H. Segur, Asymptotic solutions of the Korteweg-de Vries equation,
  {\em Studies in Appl. Math.} {\bf 57} (1976/77), no. 1, 13--44.

\bibitem{AdlervanMoerbeke} M. Adler and P. van Moerbeke, Completely integrable systems, Euclidean Lie algebras, and curves,
{\em Adv. in Math.} {\bf 38} (1980), no. 3, 267--317. 

\bibitem{BDJ} J. Baik, P. Deift, and K. Johansson, On the distribution of the length of the longest increasing subsequence of random permutations, {\em
J. Amer. Math. Soc.} {\bf 12} (1999), no. 4, 1119--1178. 

\bibitem{BaikDeiftRains} J. Baik, P. Deift, and E. Rains, A Fredholm determinant identity and the convergence of moments for random Young tableaux, {\em Commun. Math. Phys.} {\bf 223} (2001), no. 3, 627--672. 

\bibitem{BDS}
J. Baik, P. Deift, and T. Suidan, Combinatorics and random matrix theory, Graduate Studies in Mathematics {\bf 172}, American Mathematical Society, Providence, RI, 2016. xi+461 pp.

\bibitem{BPB} J. Belmonte-Beitia, V. Perez-Garcia, and V. Brazhnyi,
Solitary waves in coupled nonlinear Schr\"{o}dinger equations with spatially
inhomogeneous nonlinearities, {\em Commun. Nonlinear Sci. Numer. Simul.} {\bf 16} (2011), no. 1, 158--172.

\bibitem{BertolaCafasso} M. Bertola and M. Cafasso, Riemann-Hilbert approach to multi-time processes: the Airy and the Pearcey cases, {\em Phys. D} {\bf 241} (2012), no. 23--24, 2237--2245. 

\bibitem{BogatskiyClaeysIts}
A. Bogatskiy, T. Claeys, and A. Its, Hankel determinant and orthogonal polynomials for a Gaussian weight with a discontinuity at the edge, {\em Commun. Math. Phys.} {\bf 347} (2016), no. 1, 127--162.



\bibitem{BohigasPato2} O. Bohigas and M.P. Pato, Randomly incomplete spectra and intermediate statistics,  \textit{Phys. Rev. E} (3) {\bf 74} (2006), 036212.

\bibitem{Bohigas-deCarvalho-Pato} O. Bohigas, J.X. de Carvalho, and M. Pato, Deformations of the Tracy-Widom distribution, {\em Phys. Rev. E (3)} {\bf 79} (2009), no. 3, 031117, 6 pp.

\bibitem{Borodin} A. Borodin, Determinantal point processes, \emph{In Oxford Handbook of Random Matrix Theory}, Oxford University Press, New York (2011), 231--249.

\bibitem{BOO} A. Borodin, A. Okounkov, and G. Olshanski, Asymptotics of Plancherel measures for symmetric groups,
{\em J. Amer. Math. Soc.} {\bf 13} (2000), no. 3, 481--515. 

\bibitem{Bothner-Buckingham} T. Bothner and R. Buckingham, Large deformations of the Tracy-Widom distribution I. Non-oscillatory asymptotics, arxiv:1702.04462.

\bibitem{ChhitaJohanssonYoung}
S. Chhita, K. Johansson, and B. Young, Asymptotic domino statistics in the Aztec diamond, \emph{Ann. Appl. Prob.} \textbf{25} (2015), no. 3, 1232--1278.

\bibitem{CIK} T. Claeys, A. Its, and I. Krasovsky, Higher order analogues of the Tracy-Widom distribution and the Painlev\'e II hierarchy, {\em Comm. Pure Appl. Math.} {\bf 63} (2010), 362--412.



\bibitem{DeiftItsZhou}
P. Deift, A. Its, and X. Zhou, A Riemann-Hilbert approach to asymptotic problems arising in the theory of random matrix models, and also in the theory of integrable statistical mechanics, \emph{Ann. Math.} \textbf{278} (1997), 149--235.

\bibitem{DeiftKriecherbauerMcLaughlinVenakidesZhou}
P. Deift, T. Kriecherbauer, K. McLaughlin, S. Venakides, and X. Zhou, Strong asymptotics of orthogonal polynomials with respect to exponential weights, \emph{Comm. Pure Appl. Math.} {\bf 52} (1999), 1491--1552.

\bibitem{DeiftTrogdon} P. Deift and T. Trogdon, Universality for the Toda algorithm to compute the largest
eigenvalue of a random matrix, arxiv:1604.07384.

\bibitem{DeiftZhou}
P. Deift and X. Zhou, A steepest descent method for oscillatory Riemann-Hilbert problems, \emph{Bull. Amer. Math. Soc. (N.S.)} {\bf 26} (1992) 119--124.

\bibitem{GMT} A. Grabsch, S. Majumdar, and C. Texier, Truncated linear statistics associated with the top eigenvalues of random matrices, {\em J. Stat. Phys.} {\bf 167} (2017), no. 2, 234--259.

\bibitem{HarnadTracyWidom} J. Harnad, C.A. Tracy, and H. Widom, Hamiltonian structure of equations appearing in random matrices (Cambridge,
1992), {\em NATO Adv. Sci. Inst. Ser. B Phys.} {\bf 315} (1993), Plenum, New York, 231--245.

\bibitem{HastingsMcLeod} S.P. Hastings and J.B. McLeod,
A boundary value problem associated with the second
Painlev\'e transcendent and the Korteweg-de Vries equation,
\textit{Arch. Rational Mech. Anal.} \textbf{73}  (1980), 31--51.

\bibitem{Hone} A. Hone, Coupled Painlev\'e systems and quartic potentials, {\em J. Phys. A: Math. Gen.} {\bf 34} (2001), 2235--2245.

\bibitem{IIKS}
A. Its, A.G. Izergin, V.E. Korepin, and N.A. Slavnov, Differential equations for quantum correlation functions, \emph{In proceedings of the Conference on Yang-Baxter Equations, Conformal Invariance and Integrability in Statistical Mechanics and Field Theory}, Volume \textbf{4}, (1990) 1003--1037.

\bibitem{Johansson} K. Johansson, The arctic circle boundary and the Airy process, {\em Ann. Prob.} {\bf 33} (2005), no. 1, 1--30.


\bibitem{Johansson2}K. Johansson,
Random matrices and determinantal processes, {\em Mathematical statistical physics}, 1--55, Elsevier B. V., Amsterdam, 2006. 

\bibitem{LavancierMollerRubak}F. Lavancier, J. Moller, and E. Rubak, Determinantal point process models
and statistical inference:
Extended version, {\em J. Royal Stat. Soc.: Series B} {\bf 77} (2015), no. 4, 853--877.

\bibitem{Manakov} S. A. Manakov, On the theory of two-dimensional stationary self-focusing of electromagnetic waves, \emph{Sov. Phys. JETP} \textbf{38}
(1974) 248-253.

\bibitem{Okounkov} A. Okounkov, Random matrices and random permutations, {\em Internat. Math. Res. Notices} {\bf 2000} (2000), no. 20, 1043--1095.

\bibitem{PerretSchehr} A. Perret and G. Schehr, Near-extreme eigenvalues and the first gap of Hermitian random matrices, {\em J. Statist. Phys.} {\bf 156} (2014), no. 5, 843--876. 

\bibitem{PraehoferSpohn} M. Praehofer and H. Spohn, Scale invariance of the PNG droplet and the
Airy process,{\em J. Statist. Phys.} {\bf 108} (2002), 1076--1106.

\bibitem{Romik} D. Romik, The surprising mathematics of longest increasing subsequences, Institute of Mathematical Statistics Textbooks, Cambridge University Press, New York, 2015, xi+353 pp.

\bibitem{Sasano} Y. Sasano, Coupled Painlev\'e II systems in dimension four and the systems of type $A_4^{(1)}$, {\em Tohoku Math. J.} {\bf 58} (2007), 223--245.

\bibitem{Soshnikov2000}
A. Soshnikov, Determinantal random point fields, \emph{Russian Math. Surveys} \textbf{55} (2000) no. 5, 923--975

\bibitem{TracyWidom}
C. Tracy and H. Widom, Level-spacing distributions and the Airy kernel, \emph{Commun. Math. Phys.} \textbf{159} (1994), 151--174.


\bibitem{VMC}C. Verhoeven, M. Musette, and R. Conte,
General solution for Hamiltonians with extended cubic
and quartic potentials, {\em Theor. Math. Phys.} {\bf 134} (2003), no. 1, 148--159.


\bibitem{WBF} N. Witte, F. Bornemann, and P. Forrester,
Joint distribution of the first and second eigenvalues at the soft edge of unitary ensembles, {\em Nonlinearity} {\bf 26} (2013), no. 6, 1799--1822. 

\bibitem{XuDai} S.-X. Xu and D. Dai, Tracy-Widom distributions in critical unitary random matrix ensembles and the coupled Painlev\'e II system, arxiv:1708.06113.


\bibitem{XuZhao}
S.-X. Xu and Y.-Q. Zhao, Painlev\'e XXXIV asymptotics of orthogonal polynomials for the Gaussian weight with a jump at the edge, \emph{Studies in Appl. Math.} {\bf 127} (2011), no. 1, 67--105.


\end{thebibliography}
\end{document}